# AN ANALYSIS OF ACADEMIC PERFORMANCE
# OF UNIVERSITY STUDENTS IN NAMIBIA

A RESEARCH PROPOSAL SUBMITTED IN PARTIAL FULFILMENT
OF THE REQUIREMENTS FOR THE DEGREE OF
DOCTOR OF INFORMATION TECHNOLOGY

OF

## THE INTERNATIONAL UNIVERSITY OF MANAGEMENT
## WINDHOEK, NAMIBIA

BY

**NILS CLAUSEN, M.Sc.**

12 / 2014

THESIS SUPERVISORS:

PROFESSOR MONISH GUNAWARDANA, Ph.D.

PROFESSOR EARLE TAYLOR, Ph.D.

*"Education is the most powerful weapon which you can use to change the world."*

*Nelson Rolihlahla Mandela*

# TABLE OF CONTENTS

# TABLE OF CONTENTS

# TABLE OF CONTENTS

# I.    INTRODUCTION

## 1.    Orientation of the Study

Based on observations during my work as a lecturer for some years, its seems that a considerable proportion of university students in Namibia need to enhance their academic performance in regard to be fully competitive in a globalised working environment. This specifically includes thorough understanding of mathematics, as it forms an integral part of modern knowledge disciplines such as information technology and the management sciences. Yet students struggle or fail to engage in relevant coursework, either because of an absence of adequate material, capable educators, their own will power, or a combination thereof. This study aims to investigate the critical factors that are related to academic performance of university students in Namibia.

## 2.  Statement of the Problem

Based on data of 19,856 final marks[1] calculated and recorded within the first semester of 2014 at IUM, 13,3% of the students fail their subjects, meaning they only achieve a final mark of 49 or less out of 100. We shall define this as low academic performance.

On the other side of the spectrum, 10.4% belong to an academically high-performing group of students, achieving a final mark of 71.7 and above out of 100. We shall define academic "high performance" as the achievement of a final mark, that is situated above the standard deviation (here 15.3) of the average final mark at 56.4 out of 100. Due to the fact that every student is typically registered for more than one subject per semester, he or she can fail one and pass the others. This fact could put the student into more than one group at the same time.

The arising question is: what are the critical factors that influence academic performance of university students in Namibia, either high or low.

Subsequently two possible research problems can be derived from this question:

1. how can the number of failing students (orange-boxed, right area in figure below), from 13.3% be minimised (ideally to zero, i.e. to eliminate this area altogether)?

2. how can as many students as possible be put into the position to pass their subjects and ultimately be transferred into the area of high performance (blue-boxed, left area in figure below), i.e. scoring 71.7 or above out of 100.

---

[1] A final mark is comprised of four continuous assessments (CAs) and one exam within each semester. CAs account for 40%, examination marks account for 60% of the overall assessment of the course, i.e. the final mark.

This goal aims to widen the margin between average performance and a failing grade, which currently is only a mere 6.4 points out of a 100[2].

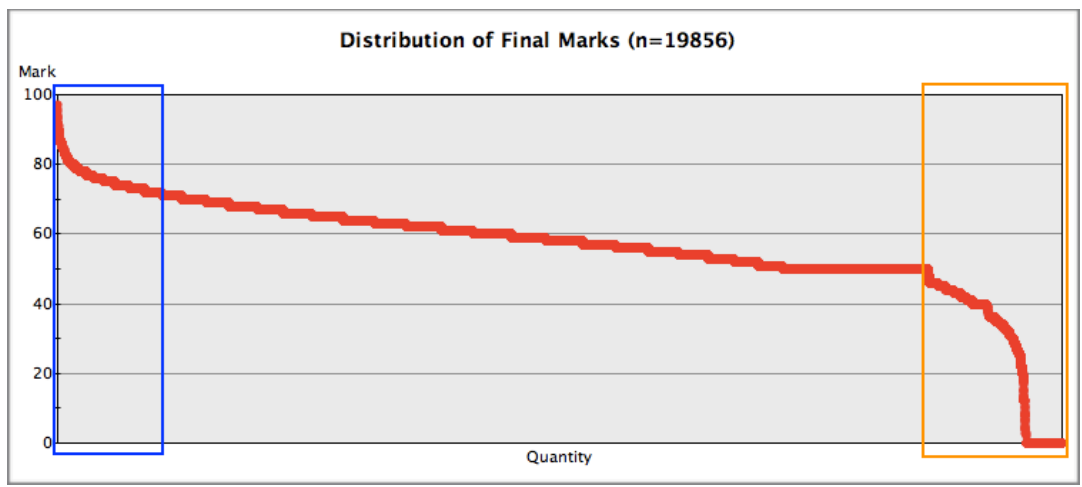

*Figure 1: distribution of final marks in the first semester of 2014 at IUM.*
*Source: iTS Student Management System, query: "distribution of final marks".*

---

[2] At the same time, current teaching and grading standards should not be weakened in favour of better-looking statistical figures. The contrary should be the case as to continuously improve grading standards by applying a high-quality, educational framework.

## 3.   Objectives of the Study

The objective of the study was to collect and analyse quantitative data in order to derive measures and actions to support the enhancement of academic performance of university students in Namibia. Socio-economical factors were not part of the analysis as reliable data could not be extracted from automated student management systems at this point in time. The literature review should cover this aspect.

The study wants lay to lay the foundation for further research, as it tries to answer the main drivers of academic performance of university students in Namibia and shows other learning institutions in Namibia, and Africa in general, ways to analyse and understand their specific situation by utilising the technical scripts and analytical methods contained herein.

This study is not meant to demonstrate the author's ability to produce more output than absolutely necessary to understand the core of the academic problem. Contrary to some widely accepted practice to include highly impressive lists of references which some authors claim to have read in full, this study only takes into account the most relevant publications from recent years. Historic views on the subject of information technology and relational databases being applied in an academic environment simply did not exist before the year 1974, when IBM began prototypically developing System R (not to be confused with the statistical software package R being used for data analysis in this study). The application of Relational Database Management Systems were used as an integral and indispensable component to capture, maintain, and analyse all data sets in this study.

A few personal words about the size of research papers I think are worth mentioning in this context. Some academics believe more is better in order show someones weight and importance in understanding a certain area of expertise. I believe, in today's fast-changing environment, less is more. Life is short and attention span even more so. If you believe the length of a research paper and it's academic value are positively associated, consider Berry, Brunner, Popescu, Shukla (2011): "Can apparent superluminal neutrino speeds be explained as a quantum weak measurement?". The answer is: "Probably not". From the author's point of view this paper illustrates that scientific relevance and the length of the research paper itself do not necessarily have to correlate.

## 4.   Hypotheses of the Study

Hypothesis I:

Intrinsic attributes of students'

- gender,

- year of birth,

- being pre- or post-independence-born (derived from year of birth)

are associated with a student's academic performance.

Hypothesis II:

Non-intrinsic attributes of students'

- chosen degree programme and

- previous school attendance

are associated with a student's academic performance.

Hypothesis III:

An application of extra-curricular learning programmes are associated with a student's academic performance.

## 5.  Significance of the Study

As stated in the Agenda 2030, Namibia aims to become an "innovative, knowledge based society; supported by a dynamic, responsive and highly effective education and training system" (Namibia Vision 2030, 2004). Implementation of Vision 2030 is formulated in each of the National Development Plans (NDPs). NDP 4 covers the time-frame between 2012 and 2017 and defines that "Namibia is characterised by a high-quality and internationally recognised education system that capacitates the population to meet current and future market demands for skills and innovation." The Namibian Government has thus clearly identified ICT[3] skills and competencies as core elements of living and participating in the 21st century and in the development of a dynamic knowledge based economy. The knowledge society is now more about skills, social networking and leading people to greater economic participation. Education has a key role to play in providing these skills and competencies.

The study is significant insofar, as it can show educators and students ways to make teaching and learning more rewarding and effective to reach these goals, which as a result will support Namibia's aim to become a knowledge-based society, create job opportunities and enable future economic growth.

---

[3] Information and Communication Technology

## 6. Limitations of the Study

As for hypotheses I and II:

1. Data from the International University of Management (IUM) from the second semester of the academic year 2013 was available and taken for analysis. IUM, as the only non-state university in the country, accounts for 22.7% of the university student population in Namibia of a total of 37,547. This could be seen as a limitation towards generalisation of the results presented in this research.

   Enrolment figures are as follows:

   | University of Namibia (UNAM) | 17,536 students[4], |
   |---|---|
   | Polytechnic of Namibia | 11,500 students (estimate), |
   | International University of Management (IUM) | 8,511 students[5]. |

2. Detail historical data is not available, as organisations in Namibia have just recently introduced centralised databases and integrated Student Management Systems.

3. Some of the attributes (database fields) are not completely or accurately filled, or might even contain wrong or misleading information. It is estimated that wrong or missing data accounts for a margin of error below 0.03, verified by various cross-checks on the data.

4. The contents of database fields might have been altered on a small scale during the querying of the database in a time-frame between 14/11/2014 and 28/11/2014, because access to the production database was done in real-time mode, and various departments could have updated single student records at any point in time. The margin of error is below 0.01, verified by various cross-checks on the data.

---

[4] Extracted from UNAM's student enrolment report 2013, http://www.unam.edu.na/wp-content/uploads/2014/07/Enrolment2013.pdf

[5] Source: database query of iTS Student Management System at IUM

As for hypothesis III:

1.  the sample size was 43 students taking the pre-test, resulting in 1,204 answers and 19 students taking the post-test, resulting in 532 answers. This could be seen as a limitation towards generalisation of the results presented in this research.

## II.  LITERATURE REVIEW

### 1.  Introduction

After the independence in Namibia in 1990, the government saw an importance in reforming the general education system. It was as one important aspect to capitalise human resource capital and to stimulate economic growth. In the beginning, the government saw the necessity of universal primary education[6] by introducing the model of Education for All (EFA). The concept became the foundation in which the Namibian education was built after independence. The structure of education was set up on four central pillars; education equity, accessibility, equality, and quality. However, regardless of the momentous investment and numerous efforts to boost education, the education system in Namibia is still seen as performing not as expected in some regards, hence it remains to be the centre of focus under the National Development Plan (NDP) No. 4 (Republic of Namibia, 2012). Educational researchers and developmental conferences have recently begun to identify the key factors behind the diminishing quality of education and particularly among university students in Namibia. The main focus of review literature concerning factors affecting education is to describe and summarise studies identifying the critical factors affecting the academic performance of university students in Namibia. The rationale of this identification is that most university students in Namibia could perform far better in their academics.

---

[6] A universal primary education is also part of the United Nations Millennium Development Goal to "ensure that by 2015, children everywhere, boys and girls alike will be able to complete a full course of primary schooling"

## 2.   Namibia as a Country

The Republic of Namibia is a country in southern Africa bordered by the Atlantic Ocean to the west and Angola and Zambia to the North, Botswana to the east and South Africa to the south east. It gained its independence from South Africa on 21st March 1990 after the Namibian war of independence. The city of Windhoek, is the largest city in the country. Namibia is a member state of the United Nations (UN), the Southern African Development Community (SADC), the African Union (AU) and the Commonwealth of Nations (CommonwealthOrg, 2014).

Namibia was inhabited by San, Damara, and Namaqua, tribes during the early days. In the 14th century the bantu tribes settled in Namibia which came to be known as bantu expansion. Later in the 19th century, Germany invaded Namibia and they were overthrown by the British and South Africans after World War I. The South Africans colonised the entire bantu and other natives of Namibia and it became an integral part of the new Union of South Africa at its creation in 1910. Namibia is the driest country in the entire African Continent and depends on groundwater. The only river is found along the national borders with South Africa, Angola, Zambia and the short border with Botswana in the Caprivi strip. In the internal parts of the country surface water is available only when the rivers get flooded after heavy rain falls[7].

Today Namibia has a population of more than 2 million people and multi-party parliamentary democracy, which is very stable. The basis of Namibian economy today is agriculture, herding and tourism. Natural resources of Namibia are mining of gem diamonds, uranium, gold, silver, and base metals. Today Namibia enjoys higher political, economic and social stability than almost any other African country (Alao, 2007).

The name Namibia is derived from the Namib Desert which is considered to be the oldest desert in the world. Before the country achieved back its independence, the area was known as German South-West Africa[8] then as South-West Africa[9], which showed the colonial occupation first by the germans and then by the South Africans on behalf of the British crown.

---

[7] Rainy season typically occurs between November to March of each year

---

[8] 1884 to 1918

[9] 1918 to 1990 (informal British- and South-African administration began in 1915)

## 3.  History of the Educational System in Pre-Independence Namibia

The people of Namibia went through all forms of prejudicial treatment on the grounds of race, gender and age during the colonial rule of germans and south africans for more than a century. Formal schooling was first introduced by european missionaries who taught the native people to read literature based on religion so that the spread of gospel[10] was made easier. This religious teaching was continued by the germans who conquered the territory in 1884 with the sole aim of colonising it by large scale immigration from their home-country, in order to expand their civilisation and dominate the native population by restricting them from areas of residences, institutions and facilities on the basis of race, so that the dominant group could maintain economic advantage over the local people. Natives were provided with limited skills in order to make sure that they remained as manual workers providing cheap labour for the immigrated population. More and more reserves were created for the natives by the germans without any adequate schools or facilities (Winschiers, 2014).

In 1914 South Africa, with the help of the British, captured the power from the germans and further intensified the colonialism which later came to be known as the creation of bantustans or homelands, and this further increased the inequality among the colonised and colonisers. The entire education system was separated on racial and ethnic lines and was designed to increase the privileges of the ruling class. Education was based to increase the skill and attitude of the oppressors to rule the native people and as an instrument of oppression. As per Fiske, Ladd (2004), in the year 1986 the colonial government spend about 3,000 South African Rand on white students, which was six times higher than the amount spent on a native student. This differential expenditure on the basis of race and class was continued until independence.

---

[10] The word gospel is a derived word from "good spell", the good news

Democracy was totally non-existent in the colonial state of Namibia and the government passed more stringent laws to increase inequalities between ethnic groups and races. They separated the schools racially by further breaking up the education into tribal schools all over the country. The government created different administration and educational systems based on race with separate schools for the white, black and coloured students. The whites were given the best education and the blacks the worst type of education. The education given to the coloured were better than the blacks but inferior to the whites. The separate education system implemented by the government was a deliberate attempt to integrate their apartheid policies so that the ruling class could maintain the status quo. In order to strengthen their apartheid policies they provided few schools which were poorly staffed and ill equipped for the non-whites due to which majority of the non-white population had to stay away from schools or compete for the limited seats in their designated schools. This was done to ensure that only a negligible number of non-whites made it to the secondary school level. These deliberately implemented education policies enabled the colonisers to suppress the non-whites into submitting to their requirements and disallowing them to become part of the society structure.

During the colonial period, the education system examinations were also conducted on a racial basis under different departments for black, coloured and white people. This isolated racial conduction of examinations was done to maintain societal differentiation of different races and to uplift the dominant classes. During the apartheid period examinations were projected as academic achievements suppressing the fact that they were conducted on gender, race, ethnicity, culture and sometimes financial categories of people. A good performance of the student did not guarantee him a high post or advance studies because the contents examined were not what the learner had written. The teacher during that time was only assigned to give marks and they were not given a chance to interact with their students or colleagues. The marks thus obtained were used as a yardstick to measure who should be admitted for higher courses (Fiske, Ladd, 2004).

Apartheid education policy was to conserve racial segregation and class distinctions to pseudo-legalise the social order. Emphasis was given to failure throughout the education system for native people. They were expected to fail and examiners were asked to set papers accordingly to ensure that large number of students other than whites would fail. This was deliberately implemented so that only a negligible number of students could process into secondary education. To put it in short: examinations were used to promote social inequality within racial and ethnic groups. The pre-independence apartheid education system had its own goals and rules were imposed to implement it, and these rules were imposed to punish the non-whites and to create more discrimination between racial groups.

## 4. Role of the Bantu Education System and its Influences on Namibia after Independence

The bantu education system was imposed on the native people of Namibia by the colonial power in the midst of resistance to apartheid. This harmful system introduced on June 7th, 1954 by South African Houses of Parliament had far reaching, long lasting social, political, economic and psychological impact on the native population. This system was developed to keep the blacks submissive to the whites and provide them with inferior education. The influence of this system still haunts Namibia and South Africa even today after so many years after the end of apartheid and independence (Sunal, Mutua, 2005). When this education system was launched the entire outcomes, purposes and contents were directly controlled by the apartheid government with the sole aim to prevent natives from receiving the appropriate education so that they do not get into aspiring positions. The hidden strategy of the system was to give the natives enough skills to serve their own people in the homelands or to work as manual labourers.

The bantu education system wreaked havoc on the black people's education and deprived them the benefits they deserved which the white people enjoyed. It was totally rejected and condemned by the natives from the time of its introduction and the devastating political, personal and economic effects of this education system can still be felt even today after all these years of independence (Fiske, Ladd, 2004). On June 26th, 1955 congress of the people issued a declaration on freedom charter in Kliptown, Johannesburg which was attended by more than 3,000 delegates from all over the country. This alliance was called the Congress of the People consisting of groups who wanted to act against the apartheid regime. It consisted of African National Congress (ANC), South African Coloured People's Organisation (SACPO), South African Indian Congress (SAIC), Congress of Democrats (white element of the Congress Movement), South African Peace Council and South African Congress of Trade Unions (SACTU). The resolution adopted in the Freedom Charter went on to state the goal of the education should be to teach the youth to love his own people and their native culture, learn to honour humans, peace, brotherhood and liberty (Lembede, Edgar, Msumza, 1996). Education should be compulsory and should be provided free and it shall be equal and universal to all children. Technical and higher education should be opened to all and scholarships and allowances should be awarded on the basis of merit.

The bantu education system implemented by the apartheid regime was devoid of all these factors taken by the congress. Even the education given to the white was designed in such a way it lacked encouragement to honour brotherhood, peace and liberty. Their education was given on the lines of christian, national education, but even lacked elements of being christian or national. The bantu education system was a devastating destruction of the dreams of those people who wanted to devote their lives for the education. Many missionaries who had devoted their life for the education were left in disgust because of the bantu education system (Edgar, 2005).

The long term effect of the bantu education system came out when a census was conducted in 1996 which is considered as a census of all people in the country. It was found that one out five black adults had received no formal education at all and only 6% had achieved education of some value. As per the census, only 25% had some sort of primary education and at the same time 33% of the population was totally unemployed. 25% of the population were earning less than 500 South African Rand per month. The bottom line is the damage created by the bantu education system is still incalculable and the effects of the atmosphere of neglect in the daily lives of Namibians and South Africans are massive. This was a dreadful and calculated crime perpetrated against them by the apartheid regime (Edgar, 2005).

After the independence in 1990 the Ministry of Education of Namibia had lots of obstacles to overcome because the pre-independence ideology of the apartheid regime had created lots of disparities and inequalities in the education sector between various ethnic groups. The pre-independence bantu education system did not meet the goals and needs of the Namibian people in regard to syllabi, teaching techniques and assessing methods. The Ministry of Education of Namibia revamped the entire education process by implementing comprehensive reforms focused on equality, quality, access, lifelong learning and democracy. A national development agenda by the Ministry of Education of Namibia embedded in the Vision 2030 was created with an emphasis on creating an innovative, knowledge based society through effective education and fully fledged training methods. Until today all ministries in Namibia are working hard to provide better education to all and produce highly trained graduates.

Today Namibia spends around 20%[11] of the national budget for education and still there is wide scope for improvement, but it has come a long way since its independence. As per statistics 95% of school age children joined schools in 2006 and the drop out ratio has come down considerably. Although considering the overall picture, despite high spending it is not producing enough output in terms of achievements and skilled work force (Sunal, Mutua, 2005). Another challenge being faced by Namibia's Ministry of Education is the recent decline in admission rates because of such factors such as lack of transportation, low population density, and children being orphaned by their parents due to HIV/ AIDS. When it comes to the standards the school curriculum quality is below average and the teaching English language to natives becomes a challenge. In order to overcome these shortcomings the Namibian government has initiated a number of reforms in the education sector. After the independence the percentage of qualified primary school teachers grew steadily in the year 1997 to 2007. In 1997 primary school teachers who were qualified to teach were a mere 21% and by 2000 it reached 36.3% and in 2007 it rose to 65.1%. The same is the case with secondary school teachers, where the percentage in 1997 was just 53% and in 2007 it rose to 90.3% (Edgar, 2005).

---

[11] This number is on par with most Scandinavian countries, being known as spending the most on a global comparison

## 5.  Namibia's Economy and Current System of Education

Namibia is categorised a upper middle income nation with a yearly average income of 64,080 Namibian $ (equivalent to 5,840 US$), and a gross domestic product of 12.58 billion US$ in 2013 according to the World Bank. However, it is also ranked as one of the most autocratic countries in the world with a Gini coefficient of 0.63 (World Bank, 2003), and its national Human Poverty Index of 28.7% in 2009, down from 37.7% in 2004 and 69.3% in 1994. The nation has made significant growth in attempting to establish an all-inclusive and equitable system of education since it gained independence. However, the poor and most marginalised children still remain to be biased within this context. This is the second report delivered to the Consultancy for Strengthening Quality of Education and Ensuring Access to Education.

The national constitution of Namibia indicates that the right to education is for all. Primary school education is compulsory, and children are supposed to remain in school until they complete primary education or until they attain the age of sixteen years. The report analyses elementary education up to grade 10 since it is the compulsory component of Namibian Education and the focus of MDG 2. According to the UNICEF (2011) report, the level in which children are accessing education is measured by Net Enrolment Rate (NER) and Gross Enrolment Rate (GER). The report indicates that there is an increase in Net Enrolment Rate and hence this shows an increase in the rate of survival in school. The increasing NER and survival rates indicate that more learners are attending school. However, the increasing GER indicate that the primary school system is becoming less efficient in term of enrolling a maximum number of children in appropriate grade with recommended age. This indicates a higher level of repetition UNICEF (2011).

Education is compulsory in Namibia for children aged 6 to 16 years. There are about 1,500 schools of which 100 of them are private schools. As per the constitution all institutions should provide free primary education. The fees for uniforms, hostels, books and school improvement have to be paid by the student's families. As per the statistics of 1997 stated a total primary net enrolment rate of 91.3%. According to the notification given out by the ministry of labor on child labor, 80% of the children aged 6 to 16 years who work continue to attend school even when they are employed. 1998 statistics shows that there were 400,325 students in primary schools and 115,237 students in secondary schools. In 2011 Namibian schools had 600,000 students of which 174,000 students were doing senior secondary education and less than 10,000 students were in the pre-primary section. If teachers in these schools are less qualified and lack specialised training due to which schools perform below expectation , dropout rates tend to become high. Further progressing, only 12% of the students go for higher education due to limited seats in universities and vocational training institutes.

In 2010 the pupil-teacher ratio in the primary section of Namibian education sector was  29.78%. The highest ratio of 33.09% was in 2004 being highest in 18 years and the lowest ratio of 29.41% in 2008. This is calculated as per the number of students enrolled in primary schools divided by the number of teachers. Detailed statistics year wise were 1992: 32%, 1998: 32.23%, 1999: 31.77%, 2000: 31.59%, 2001: 31.60%, 2003: 31.63%, 2004: 33.09%, 2005: 30.82%, 2006: 31.45%, 2008: 29.41%, 2010: 29.78%. The pupil-teacher ratio in the secondary section of Namibian education sector was 24.62% in 2007 and the highest ratio of 25.21% was in 2003 after 15 years. The lowest ratio was in 1992 with just 21.15%.

The economy of Namibia mainly dependents on the extraction and processing of minerals for export purposes. The rich alluvial deposits of diamonds make Namibia a prime source for quality gem diamonds. Namibia ranks fifth in the production of uranium in the whole world and also produces huge quantities of lead, tin, zinc, tungsten and silver. Namibia imports 50% of it's cereal requirement due to which during drought season food shortages becomes a major problem in rural areas. During the recession when the whole world was reeling under the financial crisis, the Namibian government increased revenue collection through strong tax administration, increased the export of minerals and Southern African Customs Union receipts due to which the government had enough resources to provide expansionary budgets. The main aim of the Namibian government was to sustain the economic growth a large portion of the stimulus package was spend on the education sector. In 2009/10 the education sector received 21.3% of budget allocation. Since the stimulus package was given as capital budget it was used to build schools for the primary and secondary level education. As state-owned institutions, The University of Namibia and Polytechnic of Namibia took advantage of these funds to start new courses.

## 6.  Current Educational Framework in Namibia and Comparison with Other Countries

After gaining independence in 1990, the government of Namibia placed great emphasis and resources to improve the quality of education so that the basic learning needs of all children, young people and adults were fulfilled. The desire was to provide education which was assessable, efficient, and with high quality. These desires were implemented through Article 20 of the Constitution of Namibia in 1990 which gave full priority to education as a right to free education for all the children in Namibia. This was further strengthened by the government by adopting the six Dakar goals[12] with importance to quality education of international standard, lifelong learning and democracy, childhood development at an early stage, education for girls, and for people who are marginalised and have disabilities. The reform process by the Ministry of Education was immediately followed with a standardised curriculum for secondary and primary education. The new system comprised of seven year compulsory and free education, three years of junior secondary education and two years of secondary education for children six to sixteen years.

After so many years after the independence the education sector in Namibia is still facing obstacles carried forward from the pre-independence era. Lack of funds, infrastructure, shortage of trained teachers, high failure rates are the key factors hindering the education sector. In addition to these problems the rural schools face problems like shortage of water, electricity, sanitary facilities, supply of teaching and learning materials and equipment's drop out and absenteeism rates, new challenges of education technology and HIV/AIDS.

The current programme elements of education in Namibia are manifested with the following methods.

1.  Access. Education policy on access implemented in Namibia was able to get 95% enrolment of students in the 6 to 16 years age bracket in the the last years. However access to Ovahimba and San[13] children, commercial and communal farm workers' children, and street children and HIV/AIDS patient's children is still far from being successful. The ministry has implemented specific schemes like a mobile schooling system and school feeding programmes to get and keep these children from these groups enrolled in school.

---

[12] see http://unesdoc.unesco.org/images/0012/001211/121147e.pdf

2. Equity. One of the biggest challenges being faced by the Ministry of Education in Namibia is the equal distribution of resources into all regions. This being an apartheid era legacy will take time to straighten it out. During the implementation of national development policy, efforts were made to equally distribute educational expenditure per student across the regions that lag behind and allocate funds for study materials. Introduction of new staffing norms was also implemented to bring equity to all regions. These norms were introduced in 2002 to regulate and bring equal distribution of teachers across all regions (Fiske, Ladd, 2004).

3. Quality. The quality of education comprises of many factors like qualification of the teachers, allocation of resources, and teaching materials and equipment. To overcome this hurdle the Ministry of Education implemented a large programme for provisions of class rooms, libraries and laboratories. This scheme was implemented with the help of education partners which has a long way to go to fill the back logs. As per this scheme more teachers are being trained to fill the shortage and to improve the efficiency of the existing teachers.

4. Democracy. During the apartheid period the management of the school was in the control of the teachers and even more under the autocratic rule of the principal, after independence the democratic participation of parents, students and community members in the education of their children was the focus of all schemes. Implementation of educational forums in different regions and setting up schools boards with the participation of the parents

---

[13] Ovahimba are are mainly semi-nomadic people, San are mainly hunter-gatherers

and students served as the chart for the future of the school functions and concept of democracy.

5. Curriculum. After the independence education was the major agenda of the government. Curricula were developed as per international standards along with student centred teaching methods, and assessment and semi-automatic promotion methods. New teaching materials were developed to complete the process. New areas of study like environmental studies, awareness on HIV / AIDS human rights, and lifelong learning were introduced under these schemes. A task force was set up to analyse and suggest the needs and gaps of this scheme and how to rectify them.

Some of the notably very positive developments after independence were to give top national priority to education, and in the first 10 years the 80% of adult literacy was achieved. The constitution of Namibia states that compulsory and free education should be provided to children aged 6 to 16 from grade 1 to 10. Creation of institutions like the International University of Management, community learning development centres, various colleges of education and vocational training centres have opened up opportunities for higher education and lifelong learning.

Some of the challenges faced by the government are the shortage of funds due to low economic growth and increasing budget deficits. It is believed that the new staffing norms will help in devising a formula to equally distribute the resources as per the education limits. Some constrains still exist in the education sector for the number of unqualified teachers, overcrowded class rooms, insufficient class room facilities and study materials. Some of the other constrains the education sector faces are the lack of proper utilisation of available resources, poverty, poor nutrition of students and their families, lack of sanitation, long home-to-school distances, low involvement of parents and limited fund allocation for school developments. In addition to this adds sometimes uninformed and sub-standard attitude of teachers, children, parents and the community as whole on students with special needs has become a big problem which has to be addressed at all levels.

The international community has pledged support to Namibia in their quest for education for all and the success of the Namibian education project will depend on the commitment of all potential partners and stakeholders consisting of the private sector, government of Namibia and the international community as a whole. Many stakeholders have already started taking ownership and start the developmental process of the plan. The plan is to build on the existing poverty reduction mechanism related to the education sector. The plan details the activities and strategies which are to be implemented. The implementation of the plan is time-bound and action-oriented which should be measured and realistic projections associated to the cost of all activities of the plan are made. At every stage of the implementation process resource mobilisation, advocacy, information sharing will be done with all the partners.

## 7. Influential Factors on Student Performance

In the education field it is often said that involvement of parents in their child's education produces better academic achievements. Besides parental involvement in the studies of the child there is another factor which is involved in the student's academic success and that is the socio economic status. Socio economic status of the parent can be defined as the parental income, parental education and parental occupation which correlate with the academic achievement of the child positively. As per the study by scholars, children who belongs to low socio economic family experience problems in the early school years which tend to grow as bigger problems as they grow older which in turn forces them to drop out of school. Study has also revealed that higher income family and two parent families rarely get involved in their Childs education. Lower income family headed by single parent tends you get involved more in their children's education. Involvement is also likely to occur in families where the parents are highly educated and have stable financial backgrounds. Involvement of the parents in the Childs education can have various effects academically and behaviourally. They can involve in the Childs education in many ways like reading to the child, helping with the homework, discussing the happening in the school on that day and so on.

Children who belongs to reading oriented homes and their parents are avid readers tends to score good points in reading achievement tests. The literacy skills of the child increases when the parents read it to them so that the Childs reading skills are boosted. These reading skills developed in the child can continue throughout his education. The involvement of parents in the education process can improve the performance of the child in mathematics. It can change the whole attitude of the child towards mathematics. Partnership between the parent and the school is an important factor when it comes to mathematics. It depends on how the parents socialise with their children who can affect the self-perception, ability and achievement of the child

Children's self-evaluation on their mathematic ability is closely related to what the parents perceive about them and their ability than the grades obtained. This influence of self-preservation helps them later in their career decisions. Study has also revealed that parental involvement is in some way is connected to how children perceive a particular subject and their attitude towards it. This involvement contributes a lot in the Childs achievement in science. Parents who take their child to museums and libraries tend to get more positive attitude to science.

A child's behaviour in school and social circles is closely related to the family environment and the change in behaviour may be due to neglect, passive parenting styles, and inadequate strategies to solve problems, lax disciplinary approaches and frequent conflicts at home. Schools pay little attention to partnerships and collaborations with parents and in order to improve the academic achievement teachers should form partnership with the parents. This becomes essential when the student reaches the secondary level where the parents lack confidence to assist them with their studies.

## 8. Role of the Namibia Ministry of Education

After the independence in 1990, the Namibia Ministry of Education was formed which charted out one unified education system for the nation on equitable basis. When it was launched there were lots of obstacles to overcome because of the pre-independence ideology of the apartheid regime creating numerous disparities and inequalities.

Vision 2030 stipulates compulsory and free education for children and young adults aged six to sixteen years. The level of education is classified into pre-primary and lower primary up to grades 1 to 4, upper primary up to grades 5 to 7, junior secondary up to grades 8 to 10, senior secondary up to grades 11 to 12 and university. Up to grade 4 education is provided in the native language and after that, all education is taught in English which was made compulsory. Board examination for senior secondary is conducted in grade 12 and then the senior secondary certificate is issued.

The development partners of Namibia Ministry of Education for the implementation of Vision 2030 are as follows.

1. Accenture development partnership[14]: this is a charitable institution which provides management consultancy on a global level by providing business and technology consulting services for the development sector. The aim of this institution is to channel their consulting expertise to non-profit organisations in emerging countries that do not have access to international management consultancy services. Accenture development partnership offers this service at a very economical rate that is aligned with the development sector norms. The service offered by this institution includes strategy and planning, information and communication technology, organisational development, supply chain and logistics, change management and operational effectiveness.

---

[14]   see   http://www.accenture.com/us-en/Pages/service-accenture-development-partnerships-overview.aspx

2. CECS Namibia. CECS Namibia is a non-profit organisation imparting training and support to teachers and communities in ICT[15] literacy. They mainly focus on the basics of computer literacy and once the teachers and the communities become efficient in basic skills, then an advanced literacy and proficiency in teaching computer skills is given to them (Edgar, 2005).

3. Education management information system unit. The Education management information system unit is an integral part of the Planning and Development directorate within the Ministry of Education. Their prime job is to provide statistics of the formal schooling system of the entire country and to conduct two school censuses every year.

---

[15] Information and Communication Technology

## 9. Role and Regulations of the National Council for Higher Education (NCHE)

The vision of the National Council of Higher Education is to become a leader in coordinating higher education in Namibia.

The National Council of Higher Education was formed to advise the government on all issues related to higher education. The objective of this institution is to promote and establish a higher education system, provide students access to higher education institutions, ensure quality in higher education and to advise the government on the allocation of funds to higher education institutions.

National Council of Higher Education is responsible for

- sanctioning programmes of higher education implemented in higher education institutions with the permission of the Namibia Qualifications Authority,

- monitor the quality of education given to students in higher education institution,

- taking measures to promote access of students to institution offering higher education,

- undertake research projects whenever necessary as per the instructions of the minister of education,

- advising and requesting the minister of education on the structure of the higher education system, ensuring quality promotion and assurance in higher education, allocating the funds and all aspects related to higher education

The National Council of Higher Education plays a key role in the governance and financial management of higher education and maintains a balance between equity, quality, effectiveness and efficiency in the higher education system. They are also expected to set a trend in contributing the latest methods of higher education and valuable knowledge. Since education is the key factor for growth and development in a globalised economy, it becomes a duty to ensure that there is collaboration between higher education institutions and industries so that when students a released into the labour market, they feature the necessary skills relevant to the market's requirements.

## 10. Role and Regulations of the Namibia Qualifications Authority (NQA)

The Namibia Qualifications Authority is a statutory body established as per national framework act No. 28 of 1996. The authority is made up of 29 members selected by the Minister of Education in concurrence with the Minister of Labour. NQA is mandated by legislation for the promotion of quality education and training through development and management of a flexible and comprehensive National Qualifications Framework (NQF)[16], it also promotes quality through accreditation of education and training providers and their courses in Namibia. They assist in the development of the nation by implementing systems and opportunities which allows students to develop their fullest potential without any hindrance. Namibia Qualifications Authority believes that all the students have the right to have their learning abilities recognised irrespective of when, where and how it was attained.

Namibia Qualifications Authority has the following legislative obligations for providing quality education.

- A national qualification framework to be set up and administered,

- implement an occupational standard for all jobs, posts and positions in any career structure,

- to improve the curriculum structure to achieve the occupational standards,

- promote the benchmark of all acceptable performance norms for all occupation, jobs and positions,

- give accreditation to persons and institution imparting education and training,

- encourage, evaluate and recognise skills learned outside formal education,

- act as an expert on issues related to qualifications,

---

[16] see http://www.namqa.org/framework/

- arrange facilities to collect and distribute information related to qualifications,

- ensuring that all qualifications meet the national standards,

- provide information to persons, institutions and other groups on matters related to qualifications and inform them about the national standard criteria for qualifications.

NQA itself has entered into formal Memoranda of Understanding with other quality assurance bodies thus making IUM's degrees comparable and in most cases transferrable into other legislatures. Formal agreements currently exist with:

- Botswana Training Authority (BoTA)

- Tertiary Education Council of Botswana (TEC)

- Malaysian Qualification Agency (MQA)

- Council on Higher Education (CHE)

- South Africa Qualification Authority (SAQA)

- Mauritius Qualification Authority.

## 11. Standards and Regulations of Namibian Universities

**University of Namibia**: The motto of the University of Namibia is Education, Service and Development and the mission of the University is to develop it as a leading institution in teaching and research and which will contribute to nation building by implementing the following.

- To give way to high priority research in different fields and to encourage inter-disciplinary approaches to real world problems,

- improve the standards of teaching, research, service to the community and all the functions of the university through self-improvement, constructive criticism, peer-assessment and self-evaluation,

- support the various services, skills, expertise, leadership and other facilities in the university to all people who get benefited by it regardless of gender, ethnic origin, religion, creed, physical condition, and social or economic status,

- safeguard and promote the principles of the university and provide a conducive and appropriate atmosphere and opportunity to scholars to develop their capabilities to maximum potential,

- act as a repository for the preservation, articulation, development and promotion of national culture and values of Namibian art, languages and history,

- contribute to the economic, social, cultural and political developments of Namibia by undertaking basic and applied research,

- encourage internally developed science and technology,

- provide consultancy, advisory, and extension services to promote know-how and community education throughout the country so that the socio-economic development and productivity is increased nationally to higher international standards.

The University of Namibia is the largest institution of higher education in the country. Students from all over the continent have enrolled in this diverse institution and even though the University is young, it has a student population of currently 17,536. The university consists of eight faculties and three schools under it like faculties of agriculture and natural resources, economics and management sciences, education, engineering, humanities and social sciences, law, health sciences and the school of medicine, nursing, public health and pharmacy.

Dedicated and efficient administrative professionals and the committed faculty members have elevated UNAM into the top ranks of african universities. As of today, more than 17,000 students have graduated from the University in Namibia working in different sectors and many of them holding high profile positions in the government and private sectors. UNAM welcomes students from all over the world, having the necessary entry qualifications.

Admissions start in May for postgraduates and mature age candidates and for the normal candidates it starts in June on the year of study year. All students should fill one application and mature age candidates should indicate only one course of their choice. Normal and postgraduate candidates can chose two course options as choice and the mode of study like full time or part time. All the candidates who have submitted the application will receive an acknowledge letter from the university and the successful candidates will receive a provisional admission letter from the concerned faculty or school. All other necessary requirements for the admission will be informed to the candidate from the office of the registrar.

**Polytechnic of Namibia**: The vision of polytechnic of Namibia is to become the top most institution in science and technology and to prepare new leaders for the growing economy[17]. The mission is to promote competitiveness in the national level by proving excellent education, innovation, applied research and service. The Polytechnic of Namibia contributes to wealth creation of the nation through technology oriented career courses, professional education, training, applied research and service. As an academic institution, they continuously endeavour to produce economic utility through scholarly activities by discovering, preserving and applying the knowledge (Lembede, Edgar, Msumza, 1996).

---

[17] This fact is further underlined by the planned transformation of the Polytechnic of Namibia into a Namibian University for Science and Technology (NUST). Also see: http://observer24.com.na/national/2783-new-push-for-sciences-at-polytechnic

The curriculum of this institution is to create professionals to meet the growing demands of different industries which are the driving force of the Namibian industry. In order to do this the the institution strives hard to

- make the students mature to take charge of social and economic responsibilities, problem-solving skills, integrity and humane approach to others,

- make the student knowledgeable and to develop skills to apply that knowledge in practical settings,

- train the students to become high-level specialists in all areas required for national development,

- provide facilities of the highest quality so it is available to all people who get benefited by it regardless of gender, ethnic origin, religion, physical condition, and social or economic status,

- provide monetary assistance to students who cannot afford the tuition fee at the Polytechnic of Namibia,

- safeguard and promote the principles of the academic autonomy and providing a conducive and appropriate atmosphere and opportunity to scholars to develop their capabilities to maximum potential,

- act as a repository for the preservation, articulation, development and promotion of national culture and values of Namibian art, languages and history.

- undertake applied and basic research to contribute to the economic, social, cultural and political developments of Namibia,

- improve science, technology and development to help the urban and rural communities and proving services throughout the country to improve the education as a whole,

- develop national and international unity and understanding in students,

- defend and promote excellent cultural values in the international community by encouraging self-improvement, constructive criticism, peer-assessment and self-evaluation.

All universities in the country were founded after 1990, before that time students had to go to South Africa and other countries to pursue higher education. The Polytechnic of Namibia was formed by merging Technikon Namibia and the college of out of school training as per act No. 33 of 1994 which gradually phased out vocational training courses. To get admission to courses at the Polytechnic of Namibia a student has to successfully complete grade 12 with 25 points on the evaluation scale. For international students, in addition to the general admission requirement, affected departments may ask for special admission requirements. Since the language of teaching in Polytechnic of Namibia is english, international students must be proficient in english language to get admission to a course. A minimum pass mark in grade 12 examinations will be regarded as proof of proficiency. Polytechnic of Namibia offers full-time, part-time, and distance education, and extra-curricular courses to students (Holsinger & Jacob, 2009, Winschiers, 2014) .

**The International University of Management (IUM):** IUM is a fully accredited, world-class institution determined to develop innovative knowledge workers and researchers. All programmes and degrees offered are accredited by the Namibia Qualifications Authority (NQA) within the National Qualifications Framework (NQF).

IUM is also accredited by the Education Quality Accreditation Commission (EQAC) and governed by the National Council of Higher Education (NCHE) in Namibia.

Furthermore, IUM is recognised by EU standards as an H+ (highest possible classification) tertiary institution[18].

Alliances

The International University of Management works in association with the following professional organisations and universities.

- University of Namibia (UNAM),

- Institute for Open Learning (IOL),

- Hunan Normal University (China),

- Asia e-University (Malaysia),

- The Association of Business Executives (ABE) UK,

- Bircham International University, Madrid,

- University of Technology (UTech), Jamaica,

- Shareworld Open University of Malawi.

The International University of Management was started as a non-degree awarding body and obtained the university status when it was officially launched by the president of the Republic of Namibia in 2002. IUM is located in Windhoek with campuses in Ongwediva, Swakopmund, and Walvis Bay[19].

---

[18] A full list of institutions can be retrieved from the ANABIN database, being maintained by the german Federal Ministry of Education and Research: http://anabin.kmk.org/no_cache/filter/institutionen.html

The admission to the International University of Management is done after assessment of individual applications by the university admission teams as per the admission code of practice. The university wants to treat all candidates in a fairly manner by recognising social and cultural diversity. A foreign student who applies has to get their qualifications evaluated by the Namibian Qualifications Authority and the certificate has to be attached with the application.

The admission procedures of International University of Management are as follows.

To get admission to undergraduate degree programmes a student needs 25 points in five subjects in grade 12 examination. Proficiency in english is essential for the admission. For admissions to honours degree in digital communication technology candidates should have a C pass in english and mathematics and an interview or aptitude test apart from the normal entry requirements. For an honours degree in business information management candidates apart from the normal admission procedure requires a D pass in mathematics. Candidates with 25 points in the evaluation scale with mathematics, biology and physical sciences can apply for honours degree in nursing. Mature candidates who have at least 18 points in the evaluation scale and 3 to 4 years experience social service and health sector, are eligible for application. All the applicants have to register with the Nursing Council of Namibia as nursing students before joining the course. Admissions to an honours degree in public policy management requires 25 points in the evaluation scale and a C or higher in english, mathematics and physical sciences. Admissions to honours degree in education – educational leadership, management and policy stipulates that the candidates should have a NQF level 7 or NQF level 6 qualification with 3 years experience.

---

[19] Another campus is planned in northern town of Nkurenkuru in the Kavango region, near the Angolian border.

For admissions to post graduate programmes the following criteria are required.

Candidates applying for masters in business administration should have successfully passed an undergraduate degree which is NQF accredited or evaluated at level 7 or 8. Certain professional qualifications with work experience may also be taken into account for admissions. Candidates who have successfully completed a degree or diploma in HIV/AIDS management with 65% marks and having 3 years work experience can apply for Masters in HIV/AIDS management course.

## 12. SACMEQ Report and Resources

There is a clear relationship between the quantity and quality of school infrastructure and the quality of learning in Namibia. As the SACMEQ report indicates, when classrooms are too cold or too hot, or allow in rain, learners and teachers are negatively affected. When there are inadequate toilets or lack of drinking water, learners and may feel uncomfortable to concentrate in their studies. These characteristics can ultimately also destroy the learning properties that are used to enhance learning. Likewise, when there is poor communication between the school and the surrounding community, then it becomes even harder to manage the delivery of goods and services to the school and the students (Miranda, Amadhila, Dengeinge, Shikongo, 2011).

Namibia just recently introduced a countrywide uniform testing at Grades 5 and 7 and the test has only been done once up to date. As a result, there is no local data available to show the learning outcomes in Namibian schools. However, the country participated in the Southern Africa Consortium for Measuring Educational Quality (SACMEQ) tests in 1995, 2000 and 2007. The results of the tests give general learning outcomes in Grade 6. Learners were tested in reading skills and arithmetic (Miranda, Amadhila, Dengeinge, Shikongo, 2011). Classroom observations for this report indicate mathematics in taught in elementary school classrooms lack material resources for supporting learning. These facts were confirmed during deliberations with NIED staff individuals involved in mathematics education. Upper primary classes also had inadequate materials for teaching ordinary sciences.

According to the SACMEQ report, Early Childhood Development (ECD) and pre-primary education have been extensively found to have an important effects on how children's academic performance as they proceed to higher levels. ECD and pre-primary education place the foundations in which children acquire mastery of skills at a younger which define their basic literacy. This level will determine the number of drop outs as they children progress upwards and reduce rates of repetition. If the stage is well managed, it can generate the interest of learning and going to school and even furthering studies. The quality of early childhood education ensures a smooth transition to the primary level and lays a foundation on the academic performance at primary level. Parents believe that this is the most sensitive level that defines the academic ability of a child and most parents work hard to ensure their children go through ECD (Miranda, Amadhila, Dengeinge, Shikongo, 2011). However, the major challenge in Namibia is ensuring that ECD and pre-primary education is equally accessible to the poor and marginalised group and communities that need it most.

In addition, an all-purpose or universal education is the utmost significant part of the education system in Namibia. It defines the foundation that children lay for learning as they progress to more senior levels of education. Universal education is considered to be between grades 0 up to grade 12 (Khan, 2012). The value of universal education as well as non-formal universal education is in its ability to lay a foundation for future successes. It allows successful learners to acquire skills needed for them to become competent in the labor market. They can be able to adapt to market variations and increase the volume of intake. SACMEQ's report also illustrates that the quality of universal education plays a bigger role in determining the quality of vocational or tertiary education. Vocational training is provided to train and produce skilled labor force. It also encourages a bigger number of school drop outs to join tertiary education and become productive individuals in the labor market (Nitschke, 2013). The government has put a grant system to offer loan services to teachers and graduates who are willing to teach in vocational institutions. The loan scheme is supposed to attract mainly mathematics and science teachers and graduates to teach in both vocational and secondary institutions. The quality of universal education is essential for the development of a knowledge based economy which helps the country realise its vision 2030. Namibia depends on universal education to develop self-teachable learning society.

Universal secondary education therefore provides a suitable foundation for producing labor resources needed to establish a sustainable competitive economy. With the emerging trends of information technology and globalisation of markets, secondary education improves and reinforces the capacity for continuous learning and flexible skill training. For a country to be competitive in the international market, and national competitiveness, particularly in high value added economic activities, it must have a productive system of education. Its labor force must be knowledgeable, possess skills and competencies associated with intellectual reasoning, analysis and language proficiency (Khan, 2012). Communication skills must be up to international standard and the application of information technology. Improved secondary education similarly enhances the export business sector since the sector requires the development of technical managerial skills.

Universal education also addresses the problem of inequality with regard to access of social goals. Universal education also provides an effective platform to ensure that social messages especially regarding HIV and AIDS are conveyed to the rest of the public. Moreover, secondary education provides information that in turn enhances health, reduces infant deaths, and sensitises people to adopt family planning methods. It also reduced the spread of HIV and AIDS and enhances social participation. Many and improved secondary education for girls leads to women empowerment.

Vocational training and skills development leads to economic development through their undeviating connection to labour output. Namibia's vision is to speed up economic growth by increasing labor productivity (Nitschke, 2013). The country can only achieve the goal by constantly upgrading the relevant skills required in the job market. A skilled labor force is as well fundamental to the realisation of the goals of augmented export-oriented business and value-added economic competitiveness. The goal of Namibia to shift to an economy based on knowledge will require an experienced and competitive labor force among other things. Currently, the country is experiencing scarcity of skills is one of the utmost grave constraints to employment formation and hence economic growth.

There are various ways by which tertiary education and training contribute to economic development. Tertiary education sets quality standards for the whole education system. It is focused on producing experienced technical and managerial personnel required to boost economic growth directly or indirectly. The institutions train teachers and ensure that workers and scholars have the knowledge required drive the economy. Tertiary education and training also provide business enterprises with technical support and partnership to bring about innovation based on knowledge. It provides policy experts and managers to the public and private sectors. It instills social beliefs, ideals and ethics (Nitschke, 2013). For individuals, higher education leads to higher incomes, so long as the training is in line with to the country's needs. Improved access to tertiary education by all, including disadvantaged groups, can therefore contribute to reduced poverty in absolute and relative terms. Thus improved tertiary education is essential for the achievement of Vision 2030 (Republic of Namibia, 2012).

Furthermore, a country can only solve its developmental challenges if its personnel skills in science, expertise and innovation are up to standard. Every successful economy must interconnect science, technology, innovation, and entrepreneurship using financial and strategic planning. Namibia has not given innovation the importance it deserves and this has led to a decrease in the effect of science and technology in the economy. The education sector is therefore concerned with coming up with programs that elaborates how knowledge and research can work together to improve innovation. The programs are working at consolidating vocational institutions so that they can get involved in research and development so as to make contributions on how to improve productivity and quality. The results will be improved earnings and income.

On the other hand, HIV and AIDS have posed a bigger challenge in accessing quality education. The increasing number of orphans, children caring for their younger siblings orphaned as a result of the pandemic; and children taking care of their sick parents has reduced the number of children enrolling in schools. Those in school also frequently miss lessons to take the responsibility of heading their households because they are either orphaned or their parents are terminally ill. The HIV and AIDS pandemic has remained to be the biggest threat to the growth and sustainability of Namibia's economy. Currently there is solid evidence from both inside and outside the country that education programs that teach about both HIV prevention and support programs can alleviate the effects of the virus. The Ministry of Education is receiving assistance form development partners and NGOs that are working together with the Ministry of Health and Social institutions to address the issue of HIV. They have jointly developed an effective program that is currently being implemented. The Ministry of Education has been mandated with the responsibility of ensuring that children and adults with disabilities are joined into conventional special schools. Nevertheless, such curriculums are inadequate due to unavailability of special schools and lack of trained teachers who understand and cater for the challenges experienced by these learners.

The SACMEQ report also elaborates that knowledge, expertise, and technology have now become critical factors of economic growth as compared to natural factors of production such as land, capital, and labor. The previous three decades has seen production gradually become more knowledge demanding by investing research and development, information technology, product strategy, engineering, and quality measurement. It is also investing in analysis, training; sales and marketing and management. This is so as to improve the production of goods and services. The knowledge demand of production has gradually stretched beyond the extraordinary technology sectors to reform a wide range of outdated industries.

## 13. Self-regulated Learning and Education Achievement

Educational practitioners have in recent times started to identify key processes that university students use to self-regulate their academic learning. Self-regulated learning means that students begin to take responsibility of their own learning. For example, Zimmerman, Schunk (2001) illustrate Benjamin Franklin's autobiography. He exclusively described the methods he used to improve his learning, knowledge and self-control. He worked hard to enhance his writing skills by choosing to read archetypal written models and endeavouring to emulate the authors. He defines how established learning objectives for himself, indicating his everyday progress in a record book. He taught himself how to write and he testifies that the procedure moulded him into a better individual by improving his memory and cognitive ability.

Self-regulated learning has provided a significant view on current psychological research about academic learning. According to Zimmerman, Schunk (2001), self-regulated learning is the ability of students to use their intellectual and metacognitive approaches to regulate their learning. The importance of self-initiative in learning has been reaffirmed by present-day influential leaders such as Gardner, the former United States Secretary of Health, Education, and Welfare. He suggested that the final objective of the education system is to transfer to individuals the burden of pursuing their own learning. Up to now, experts have come up with little experimental evidence to show that students have become masters of their own education (Zimmerman, Schunk, 2001). However, the last few years have seen researchers attempting to identify and study some of the key processes that students use to acquire their academic knowledge. Apart from the perspective of self-regulate learning being unique, it equally has implications on the manner in which teachers should interact with student and the way the school should be organised.

According to Pintrich (1999), self-regulated students have common characteristics that define their competence. They approach educational tasks with self-confidence, resourcefulness, and meticulousness. They are aware when they know a fact or possess a skill and when they do not. They are proactive and unlike their passive classmates, they go out there to seek information when needed and take the necessary steps to master it. When they encounter obstacles such as inadequate study conditions, unclear teachers, or complex books, they find a way to succeed. There are three features that are involved in the definition of students' self-regulated learning: first, it is important to differentiate between self-regulation such as perceptions of self-efficiency and methods put in place to improve these processes when defining self-regulated learning. As Zimmerman, Schunk (2001) state, strategies of self-regulated learning are actions and process used by learners to acquire information or competencies that involve action, purpose, and influential perceptions. As much as all students initiate their own learning to some extent, self-regulated students are different. The difference is that self-regulated students are aware of strategic relationship between self-initiated learning and learning outcomes. They also know how to use these strategies to achieve academic excellence.

The second feature used to define self-regulatory learning is a cycle of self-oriented response or feedback. The cycle is a repeated process where learners track the efficiency of their learning methods and then respond to the feedback in several ways. Their reactions involve secretly changing their personal perceptions and openly changing their learning behaviour. Changes in behaviour involve changing the usual learning methods while changes in self-perception include self-esteem and self-concepts (Sun, Tsai, Finger, Chen, Yeh, 2008). Different theories of self-initiated learning have different descriptions about the feedback loop. For instance, operant theories drawn by Pintrich (1999) claim that feedback involve openly changing learning behaviour by individual learning, individual recording, and individual support. Social cognitive theorists such as Bandura (1989) caution that the loop should not just be perceived in terms of negative feedback, that learners report positive responses as well. Irrespective, of the contrasting theories regarding what is monitored and the manner in which results are interpreted, nearly all scholars assume that self-regulated learning is based on ongoing feedback on the efficiency of learning.

The third feature of defining self-initiated learning is a hint of how and why learners choose to use a particular method of learning. Since self-regulated learning is made up of defined methods and responses of learning, students require ample preparation time, caution, and strength to initiate and control them well. Students will only be motivated to initiate self-regulation if the outcomes of learning are going to improve their academic performance. Operant theorists suggest that the learners can only initiate their own learning if they receive rewards or punishments such as being socially endorsed enhanced status, or material benefits. Phenomenological philosophers on the other hand assume that learners are motivated if their self-esteem is boosted or they reach self-actualisation (Zimmerman, Schunk, 2001).

Students who practice self-regulated learning have adopted strategies that help them in achieving academic excellence. The strategies are based on feedback about the effectiveness of learning and skill. Both laboratory and field research have been carried out to find out the benefits of self-controlled strategies in achieving academic excellence. Researchers include Sun, Tsai, Finger, Chen, Yeh, (2008), and Zimmerman, Schunk (1999). They suggest that quite a number of metacognitive, motivational, and behavioural strategies have been studied in a many universities and laboratories worldwide. There are three overall classifications of self-regulated learning approaches or strategies: First are cognitive learning approaches, self-initiated approaches to regulate reasoning, and lastly, resource management approaches.

In reference to cognitive learning approaches, Zimmerman, Schunk (2004) describe how self-regulated learners apply this strategy. They state that the approach is characterised by practice or rehearsal, explanation, and organisational skills. They are significant reasoning strategies that improve academic performance in the classroom. These approaches can be used in simple tasks that require rote learning such as remembering information, verses or lists. It can also be used in more multifaceted tasks that are involved with deep understanding of the information.

Pintrich (1999) explains that practice or rehearsal approach is where self-regulated students recite items that they are yet to learn or recite words aloud as they read their books. It is accompanied by highlighting or underscoring texts that are considered relevant to keep in memory. The rehearsal approaches help learners to read and choose helpful information from books and other manuscripts and then store the information in their memory. However, the strategies do not display an actual level of cognitive processing. Explanation strategies comprise of paraphrasing or summarising the material to be learned by creating similarities, reproductive note-taking. Students reorganise and combine ideas in their notes, different from just writing down straight notes directly from the material. They are able to explain and elaborate to others the ideas in the material to be studied and also answer any questions asked (Pintrich, 1999). Organisational strategy on the other hand is an equally profound processing approach. It involves highlighting the main idea from a text and outlining the specific material to be learned. They then use specific techniques for selecting and organising the ideas in the material, such as sketching a plan or a network of important ideas and identifying the flow of ideas or structure of a manuscript.

Apart from cognitive learning strategies, students also use metacognitive strategies to improve their academic performance. There are two universal features of metacognition; the understanding of cognition and self-control of cognition (Brown, Bransford, Ferrara, Campione, 1983). There has been some philosophical and experimental confusion regarding the meaning of metacognition. The confusion is as a result of confusing issues of the knowledge of metacognition and the control of metacognition (Brown, Bransford, Ferrara, Campione, 1983). Pintrich (1999) has recommended that the limitation of metacognitive knowledge among students allows for self-regulation, where by students can monitor and control their own reasoning activities and behaviour. The research program conducted at the University of Michigan focused on the strategies that students use to design, monitor, and regulate their reasoning, and not their metacognitive knowledge. According to Pintrich (1999) and Zimmerman, Schunk (2001), there are three types of metacognitive strategies which are planning, monitoring, and regulating. Even though these three types of strategies are extremely related theoretically and also appear to be extremely interrelated empirically, they can be debated distinctly.

Investigations carried out in various studies indicate that planning activities involve setting goals for learning, scanning a script before reading, writing down questions before reading a script, and analysing the task in a problem. These planning actions are supposed to help the student plan how to use their cognitive strategies and also help them to trigger or prime relevant aspects of prior knowledge, making the organisation and comprehension of the material much easier.

Zimmerman, Schunk (2001) suggest that students should monitor their own reasoning and academic behaviour so as to be self-regulated. In order to be self-regulating, students must set goals or standards that can be used to compare the performance of activities against the set goals. Monitoring activities include tracking of attention while reading a text or listening to a lecture, self-testing through the use of questions about the text material to check for understanding, monitoring how one understands a lecture, and using the examination approach such as tracking speed against available time. Regulation strategies on the other hand are almost related to monitoring strategies. As students track their learning and performance against some goal or standard, regulation comes in to play a role. Regulation strategy suggests that there is a need to align behaviour with the set goal or standard. For example, students can ask themselves questions in the process of reading and provide answers so as to monitor their understanding. Afterwards, they go back and revisit a section of the text. This process of revisiting text is known as regulatory strategy. A different type of self-regulatory strategy for reading happens when a student slows the pace of their reading when they come across more difficult or unfamiliar text. For example, revising any type of course material such as lecture notes, texts, lab notes, earlier examination papers, that one does not remember or understand that well while studying for an examination. During a test, skipping questions and returning to them later is another strategy that students can use to regulate their behaviour. All these strategies are assumed to improve learning by helping students correct their studying behaviour and repair deficits in their understanding.

According to Pintrich (1999) and Zimmerman, Schunk (2001), another component of learning and self-regulatory strategies, is the resource of management strategies. It concerns strategies that students use to manage and control their environment. Examples include managing and controlling their time, their effort, their study environment, and other people, including teachers and peers, by seeking help from others. They suggest that the resource management approaches are assumed to help students adapt to their learning environment as well as change the environment to fit their goals and needs.

There are many prototypes of motivation that are appropriate to improve student learning (Pintrich, 1999). They include beliefs of self-worth, where students judge their own capabilities in attaining academic success. Secondly, are the beliefs of the value of academic excellence. Students perceive the importance of studying and their interest in learning. Lastly, goal orientation explains whether students are focused on mastery and learning of the task, and external reasons for doing the task.

The relationship between these three general types of motivational beliefs has been studied. The use of self-control approaches in both middle school settings and university settings has also been explored. The research program has focused on describing how different motivational beliefs help to support, sustain, or expedite individual initiated learning. The research has largely been related with data collected over a long period time. In the studies of middle school children, one group of students has been followed over the course of three years in school (seventh, eighth, and ninth grade) with five waves of data collected; other middle school studies have sampled fewer waves of data. In all, data on approximately 1000 middle school students have been collected over the years. In the college studies, data have been collected mainly at two points in time, early in a 15 week semester and then again late in that same semester. The college samples have included over 3,000 cases in the different studies.

All the research has used the Motivated Strategies for Learning Questionnaire (MSLQ) (Pintrich, 1999), a report tool intended to measure students' motivation and self-regulated learning in classroom situations. Data have also been collected on the students' performances on examinations and papers, as well as their final grade for the course. It is important to note that all of these studies have been conducted in classrooms level. Studies in the middle school sampled courses such as arithmetic, Science, English or Social Studies. At the college level, studies sampled Preliminary Psychology, English Literature, Preliminary Biology, Preliminary Sociology, and Calculus. Students' motivation and self-regulated learning are assumed to be specific to learning situations and thus, a focus on the class or course level was seen as the most appropriate level of context. Consequently, the MSLQ is not designed to assess students' global motivation and self-regulation for school or college, nor is it sensitive enough to detect differences in motivation or self-regulation as a function of different tasks within a class. For example, a multiple choice test versus a research paper within the same class.

## 14. E-Learning and Academic Success

E-learning is the utilisation of information and telecommunication technologies (ICT) to convey educational information and training. The tremendous growth of information and communication technology has seen an emergence of e-learning being the model of modern education. Manochehr (2006) and Ellram, Easton (2006) outline the advantages of e-learning, such as improving interactions between students and teachers, and between students themselves. A lot of research conducted regarding this subject suggests that the attitude of students towards computers or information technology is a significant factor in e-learning satisfaction. Learner attitude can be defined as the student idea to take part in computer networks. For instance, a more positive attitude towards e-learning means that students are not scared of understanding the complexity of computers. Therefore, it will result in more satisfied and effective learners in an e-learning environment. In addition, Marakas, Johnson, Clay (2007) denote that attitude influences the interest to learn. Positive attitudes towards computers increase the chances of successful computer learning while negative attitudes reduce learning. Hence, the above research is a confirmation that students' attitude towards information technology is an essential factor in determining academic performance (Ellram, Easton, 2006) d'Oliveira, Carson, James, Lazarus (2010) show that anxiety when using a computer significantly affects the satisfaction of e-learning among students. Computers are the central media machines that are used to convey information; hence incompetency in computer skills can affect e-learning satisfaction. There are two types of computer anxiety which are caused by mental pressure. Trait anxiety is manageable and an enduring internal characteristic while state anxiety is caused by the external environment. Preceding research has shown that computer anxiety is comprised of more of state anxiety. It is an emotional fear of possible negative results such as damaging the equipment or looking foolish. If computer anxiety is higher, it will result in lower level of learning satisfaction. Lower levels of learning satisfaction in turn results in poor academic performance. Manochehr (2006) also proposes that computer anxiety interferes with individual attitudes and behaviour.

In the view of Marakas, Johnson, Clay, Paul (2007), self-efficacy is a person's predisposition toward a specific functional characteristic. It is where an individual evaluates the possible effects and the likelihood of achieving success before carrying out a task. Students who possess high levels of self-efficacy are more confident in attaining e-learning activities and enhancing their learning satisfaction. Various studies have explored how self-efficacy influences the ability of students to recognise effects of e-learning. Manochehr (2006) indicates that self-efficacy is a significant factor in foretelling about the possible effects taking part in a computer network-based learning. d'Oliveira, Carson, James, Lazarus, (2010) also specify that particular self-efficacies inclined to the internet meaningfully influence results whenever students are searching for items or keywords on the internet. They point a research done on 122 learners. The research concluded that students with greater self-efficacy are more likely to embrace e-learning and in turn significantly excel in their examinations. The research defines online self-efficacy as students' capacity to assess their ability to use the Internet to carry out activities linked to e-learning.

Research conducted in earlier times indicates that the speed at which teachers or instructors respond to students influence students' satisfaction. This is according to d'Oliveira, Carson, James, Lazarus (2010). The whole idea is that when students are confronted with problems when the using the internet for their studies, the instructor should offer them with immediate assistance to encourage them to continue with learning. They illustrate that if teachers fail to respond to students on time, their learning will be negatively influenced. Therefore, if a teacher is proficient in supervising e-learning activities and responding to the needs and problems of students needs promptly, learning satisfaction will improve.

A majority of researchers savvy in the realm of education recommended a social influence model of technology which states that the attitude of group members or that of the supervisor regarding technology has a negative influence on students' perceptions. Psychology explains that individuals get to learn and develop their own synchronised patterns by observing and imitating the people around them. They imitate their actions, attitudes, and emotions. Marakas, Johnson, Clay, Paul, (2007) point out that teacher's approach towards e-learning and information and communication technologies influence the outcomes of e-learning. It is because teachers are primary actors in educational activities. It is highly suggested that the attitudes of teachers towards e-learning should be dealt with in system evaluation so as to expound on the behaviours of online course users explicitly.

E-learning courses are flexible in terms of time, location, and strategies; hence it is easier to facilitate participation and satisfaction of learners. d'Oliveira, Carson, James, Lazarus (2010) provide a deeper insight on this fact. In addition, removal of barriers allows for more active interaction that promotes formation of constructive learning and openings for supportive learning. Students can communicate promptly, anytime, anywhere in the absence of time and space restrictions. Moreover, the fact that learning is computer-generated eliminates clumsiness related to face-to-face communication in traditional classrooms. Learners can express their thoughts without reticence and ask questions through discussion. Presently, most e-learning courses are free of charge in applicable learning and continued education programs, and learners are mostly people who are employed (Ellram, Easton, 2006).

Students consider the quality of well-made e-learning programs as an important factor when choosing to take part in such programmes. Quality is an additional important factor influencing learning effects and satisfaction in e-learning. Under the productive or supportive learning ideal, interactive communications and media presentation provided by information and communication technologies can help learners develop high-level thinking models and establish conceptual knowledge. The virtual characteristics of e-learning, including online interactive discussion and brainstorming, multimedia presentation for course materials, and management of learning processes, assist learners in establishing learning models effectively and motivating continuous online learning. Therefore, the quality of e-learning courses is also considered a significant factor in learner satisfaction (Manochehr, 2006).

## 15. Commonwealth of Learning and Education Achievement

The Commonwealth of Learning (COL) supports the expansion of Open Distance Learning (ODL) so as to meet the increasing need for primary, secondary and even higher education in Namibia. COL came up with a unique initiative which was to improve and reinforce open schools. It therefore aided the creation of Commonwealth Open School Association (COMOSA) which is comprised of partners from all commonwealth countries. In addition, COL's Graduate Diploma in Legislative Drafting builds professional skills through ODL (UNICEF, 2011).

The Commonwealth of Learning (COL) adjusts e-learning methods of education to be in line with the required capacity improvement needs of organisations around the globe. COL improves and provides effective training resolutions in corporation with the United Nations (UN) and other global agencies; this was supported by the study by Nitschke (2013) on vocational trainers in Africa, and Namibia in specific. The partnership work is executed on the terms of first paying for services with full cost recovery. The topics covered in the course include effective communication, skills in report writing, operational data management and debt management. On the other hand, COL's two-yearly Pan-Commonwealth Forum on Open Learning (PCF) is the number one conference in the world that advocates for learning and development. It brings together people standing in for educational institutions, governments and development organisations to explore issues related to Open Distance Learning and development. More than six hundred delegates from seventy nations attended the conference in Kerala, India in 2010 (UNICEF, 2011).

## 16. Conclusion of Literature Review

The concept of Education for All (EFA) became the foundation in which the Namibian education was built after independence. The system of education was founded on four main pillars; education equity, accessibility, democracy, and quality. However, despite the momentous investment and numerous efforts to strengthen education, Namibia's system of education is still seen as performing below its potential and, hence it remains a strategic sector under the NDP 4. The primary focus of this literature review describes and summarises the studies identifying the critical factors affecting the academic performance of university students in Namibia. Some of those factors are associated with the economy of the country. Namibia is ranked as one of the most autocratic countries in the world with a Gini coefficient of 0.63 (World Bank, 2003), and its national Human Poverty Index of 28.7% in 2009, down from 37.7% in 2004 and 69.3% in 1994. As a result, the poor and most marginalised children still remain to be biased when it comes to accessibility of education. The national constitution of Namibia indicates that the right to education is for all. Primary school education is compulsory.

Among the critical factors affecting academic performance in Namibian Universities is self-regulated learning. It means that students begin to take responsibility of their own learning. According to Zimmerman, Schunk (1999), self-regulated learning is the ability of students to use their intellectual and metacognitive approaches to regulate their learning. Students who practice self-regulated learning have adopted strategies that help them in achieving academic excellence. Researchers include Tsai, Finger, Chen, Yeh, (2008), and Zimmerman, Schunk (1999). They suggest that there are three overall classifications of self-regulated learning approaches or strategies: first are cognitive learning approaches, self-initiated approaches to regulate reasoning, and lastly, resource management approaches. Another factor is e-learning, which is the utilisation of information and telecommunication technologies to convey educational information and training. Manochehr, (2006) and Ellram, Easton (2006), outline the advantages of e-Learning such as improving interactions between students and teachers, and between students themselves. Students' attitude towards information technology is an essential factor in determining academic performance.

According to the SACMEQ report, Early Childhood Development (ECD) and pre-primary education have been extensively found to have an important effects on how children's academic performance as they proceed to higher levels. In addition, an all-purpose or universal education is the utmost significant part of the education system in Namibia (Light, Pierson, 2014). It defines the foundation that children lay for learning as they progress to more senior levels of education. Universal secondary education therefore provides a suitable foundation for producing labor resources needed to establish a sustainable competitive economy. On the other hand, HIV and AIDS have also posed a bigger challenge in accessing quality education. The increasing number of orphans, children caring for their younger siblings orphaned as a result of the pandemic; and children taking care of their sick parents has reduced the number of children enrolling in schools.

## III. METHOLODOGY

### 1.  Research Design

A quantitative research design is chosen to examine associations, possible correlations, or covariations between variables in the area of this research.

### 2.  Research Ethics

Detail data from study participants will not be disclosed on an individual data level at any time, data is only disclosed on an aggregated and therefore anonymised state. All data is kept in safe environments using appropriate and up-to-date data-encryption techniques.

A peer-review process is applied to assure quality standards and to ascertain the originality of the research.

Organisational and individual consents were granted in writing by those participating in the study. Opting out of the study was possible for the organisations and individuals at any point in time. All original material and corresponding electronic communication, such as e-mails, are archived and stored for at least 10 years to support confirmability and traceability. All raw data used is stored in electronic format on a file server and can be downloaded at any time for further analysis.

### 3.  Population

Research findings are generalised to the whole population of university students in Namibia, regardless of their origin and nationality, enrolled at tertiary learning institutions University of Namibia, Polytechnic of Namibia and the International University of Management. By the end of 2013, the total number of university students was 37,547.

## 4. Samples

Data samples were drawn from IUM's internal Student Management System "iTS iEnabler" and from paper-based tests performed by students from IUM.

iTS runs on a centralised Oracle Relational Database Management System that was queried using SQL statements, all data sets were accessible during the time of querying the database. This means 100% of the data in the system was available for analysis, comprising 22.7% of the student population in Namibia.

Tests were performed in exam-compatible environments with invigilation allowing no further tools such as calculators, mobile phones, and written material.

## 5. Research Instruments

Research instruments included two paper-based tests being taken from 43 (pre-test) and 19 students (post-test) in a time-frame between 28/10 and 27/11/2014, transcribed into an electronic file format between 28/11 and 30/11/2014 and the querying and various snap-shootings of the central database in time frame between 14/11 and 28/11/2014.

## 6. Procedure

For retrieval of the student mark and exam data, preparative work for this research started in May 2009 with a feasibility study for the implementation of an integrated and centralised Student Management System at the University of Management to facilitate analysis on student data. At that point in time, all student records were recorded and archived in paper-based and electronic but decentralised spreadsheet formats in various locations of the different campuses in Swakopmund, Walvis Bay, Ongwediva and Windhoek. Following the positive outcome of the feasibility study, a technical and organisational concept was authored to support the implementation of such system until the end of 2013. Various tasks in these concepts had to be completed to finally reach the decision to conduct a market research on products being able to meet all requirements. A range of suppliers were being invited to show their solutions. In 2012 IUM's management decided to purchase and implement iTS iEnabler, a solution widely being implemented at universities in Botswana, South Africa, Swaziland, and Namibia[20].

At the end of 2013 iTS was readily implemented and able to be used on a day-to-day operational basis. At the beginning of the academic year 2014, all master data of enrolled students were being transcribed and transferred into the system, new students were registered with the new iTS application. During the the first semester of the 2014 academic year, all student marks were directly input by the lecturers and validated by the registrar with frequent data check routines. The system therefore laid the necessary foundation to enable extensive and detail analysis of student data.

Preparation for the paper-based tests started in the second semester of the academic year 2014, i.e. in July 2014. Four different test types were designed, contents were derived from the California High-School Exit Examination[21] (CAHSEE), that acted as pre- (document no. S00173 and S00175) and post-tests (document no. S00177 and S00179) to determine student's skill levels in mathematics. The study was announced on IUM's whiteboards in mid-October 2014, all registered undergraduate students were also invited by e-mail sent to their campus e-mail addresses shortly before the initial presentation and the pre-tests of the study to students on 28/10/2014. The key points of the initial presentation were as follows:

1. Stating the motivation: new technologies replace or complement existing ones; technical knowledge and skills are very important personal assets today; and even more so tomorrow; Namibian students need support to be prepared to understand and shape innovation; smartphones and social media are mostly used for entertainment and communication, but less for education.

---

[20] see interactive map representation of data at https://mapsengine.google.com/map/edit?mid=z7t3_8C-lGfA.kvRq6AIPYLO8, sources: Integrated Tertiary Software (Pty) company profile, press releases and own research

[21] http://cahsee.cde.ca.gov

2. Explaining the e-learning initiative at IUM: new technologies can only be mastered with the right skills; the source of these skills do not originate from a singular source; e-learning connects you to the best lecturers at the best universities; universities will change their shape: a home to their students with more intense class-room lectures and complemented by compatible sources from other institutions; learning needs to happen at places and times fostering optimal flow: making ability and willingness coincide.

3. Detailing the research study: a standardised math test with 28 questions is taken by undergraduate students; test takers will be split into two groups: one group is supplied with traditional reading material, the other group is supplied with an e-learning curriculum to be performed on their laptops or smartphones; after a short learning period, another standardised math test with 28 similar questions is taken by both groups; results will be analysed for statistical significance to determine whether e-learning offers a comparative advantage over traditional learning methods; following the presentation, students can take part in the math test, containing 28 questions; duration will be 60 minutes or less; results will not influence any grades at IUM; personal data will not be published or made available to other sources; shortly after the test, students will receive detailed instructions by e-mail on how to prepare for the second test; after two weeks, the second test will be performed; students will help to influence future curricula at IUM.

All test answers were subsequently input into an electronic spreadsheet file to enable analysis of the data and students were informed of their individual results by e-mail.

## 7.  Data analysis

Data extraction was done with DbVisualizer 6.5.12 and Oracle JDBC connection drivers to an Oracle RDBMS instance on Version 11g. Resulting data sets were analysed with R 3.1.2, R Studio 0.98.1091, and Wolfram Mathematica 9. Cross-checking of results was done using the online version of R Studio[22]. Graphs and charts were created using DbVisualizer, R with R Studio, and Wolfram Mathematica in their above stated versions. Google Map Maker was used for geo-coding and mapping, results are in on-line retrievable format from Google Maps. Authoring this document was done with Apple Pages 5.5. All front-end software applications were executed on an Apple MacBook running OS X Yosemite 10.10.1.

Data was analysed using averages, standard deviations, percentages, linear regressions with correlation coefficients, and two-sided significance tests with a one-on-one cross-comparison of all available attributes. Multiple combinations of attributes were not analysed as this leads to an exponential increase of calculations with limited applicability of results.

---

[22] http://glimmer.rstudio.com/odnl/ab-test-calculator/

# IV. FINDINGS AND CONCLUSION

## 1. Intrinsic Attributes and Academic Performance

**Association with gender:**

|  | Female | Male |
|---|---|---|
| **Number of high performers** | 1,411 | 659 |
| **Number of low performers** | 1,620 | 1,026 |
| **Total number of final marks** | 13,049 | 6,847 |
| **Average final mark** | 56.8 out of 100 | 55.6 out of 100 |
| **Standard deviation of final mark** | 15.0 | 15.7 |
| **Percentage of high performers** | **10.8%** | 9.6% |
| **Percentage of low performers** | 12.4% | **15.0%** |

*Data table 1: final marks by gender, queries used "distribution of final marks by gender (female)" and "distribution of final marks by gender (male)".*

**Findings:**

1. Average final marks do not differ significantly between male and female students, therefore showing a weak association.

2. With a confidence level of 99% in a two-sided significance test, female students have a significantly higher chance than male students to belong to the high-performing group of students.

3. With a confidence level 99.9% in a two-sided significance test, male students have a significantly higher chance than female students to belong to the low-performing group of students.

**Association with year of birth:**

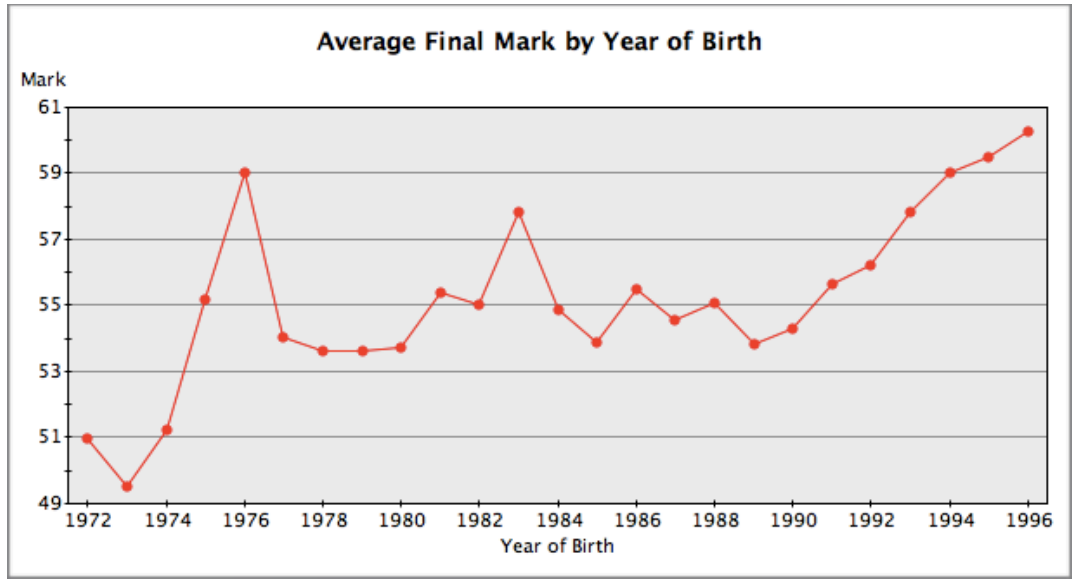

*Figure 2: average final marks by year of birth.*

*Source: iTS Student Management System, query: "average final marks by year of birth".*

| Year of Birth | Average Final Mark | No. of Marks |
|---|---|---|
| 1972 | 51.0 | 34 |
| 1973 | 49.5 | 55 |
| 1974 | 51.2 | 52 |
| 1975 | 54.7 | 74 |
| 1976 | 59.0 | 69 |
| 1977 | 54.1 | 105 |
| 1978 | 53.6 | 108 |
| 1979 | 53.6 | 120 |
| 1980 | 53.7 | 185 |
| 1981 | 55.4 | 148 |
| 1982 | 55.0 | 188 |
| 1983 | 57.9 | 288 |
| 1984 | 54.9 | 300 |
| 1985 | 53.9 | 362 |
| 1986 | 55.5 | 536 |
| 1987 | 54.5 | 632 |
| 1988 | 55.0 | 914 |
| 1989 | 53.8 | 1,348 |
| 1990 | 54.3 | 2,088 |
| 1991 | 55.7 | 2,374 |
| 1992 | 56.2 | 2,980 |
| 1993 | 57.8 | 3,027 |
| 1994 | 59.0 | 2,271 |
| 1995 | 59.5 | 832 |
| 1996 | 60.3 | 95 |

*Data table 2: average final marks by year of birth.*

*Source: iTS Student Management System, query: "average final marks by year of birth".*

**Findings:**

1. By performing a visual estimate, the numbers indicate that younger students tend to score higher on average than older students (with an exception of students born in 1976 and 1983), with a gap around the year 1989 and followed by an upstream trend.

2. A linear regression analysis using the above data set for years of birth 1972 - 1996 supports the assumption of a positive correlation (younger students score higher on average) with b = 0.2464 and thus y = 0.2464$x$ - 433.7834 and a correlation coefficient r = 0.6855.

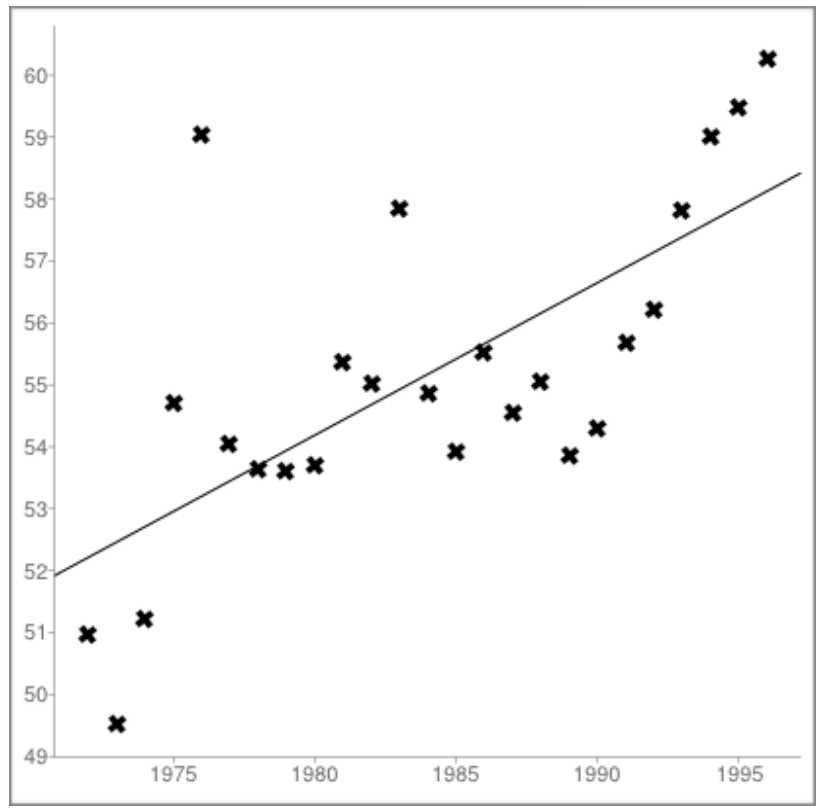

*Figure 3: average final marks by year of birth with linear regression slope.*
*Source: iTS Student Management System, query: "average final marks by year of birth".*

**Association with being born pre- or post-independent (born before and on/after the year 1990):**

|  | Pre-Independence Born | Post-Independence Born |
|---|---|---|
| **Number of high performers** | 531 | 1,430 |
| **Number of low performers** | 922 | 1,682 |
| **Total number of final marks** | 5,695 | 13,654 |
| **Average final mark** | 54.7 out of 100 | 56.9 out of 100 |
| **Standard deviation of final mark** | 16.8 | 14.6 |
| **Percentage of high performers** | 9.3% | **10.5%** |
| **Percentage of low performers** | **16.2%** | 12.3% |

*Data table 3: final marks by gender, queries used "distribution of final marks (pre-independence born)" and "distribution of final marks (post-independence born)".*

**Findings:**

1. Average final marks do not differ significantly between both groups.

2. With a confidence level of 99% in a two-sided significance test, post-independence born students have a significantly higher chance than pre-independence born students to belong to the high-performing group of students.

3. With a confidence level 99.9% in a two-sided significance test, pre-independence born students have a significantly higher chance than post-independence students to belong to the low-performing group of students.

4. A linear regression analysis using the above data set for years of birth 1990 - 1996 supports the assumption of a even stronger positive correlation (students born after independence score higher on average) with b = 1.010 and thus y = 1.010$x$ - 1,956.1677 and a correlation coefficient r = 0.9900.

## 5. Non-Intrinsic Attributes and Academic Performance

**Association with degree programme:**

| Rank | Degree Programme | Average Final Mark | Number of Final Marks |
|---|---|---|---|
| 1 | **Masters Business Administration and Management** | **66.7** | 79 |
| 2 | **Honours Bachelor Degree Nursing** | **66.7** | 113 |
| 3 | **Higher Certificate HIV and AIDS Management** | **66.4** | 577 |
| 4 | **Masters Business Administration and Finance** | **66.3** | 34 |
| 5 | **Masters in Business Administration and Human Resource Management** | **65.4** | 36 |
| 6 | **Certificate German** | **63.4** | 106 |
| 7 | **Honours Bachelor Degree in Education** | **63.4** | 74 |
| 8 | **Higher Certificate Marketing Management** | **61.7** | 409 |
| 9 | **Honours Degree Human Resource Development and Management** | **61.2** | 669 |
| 10 | **Higher Certificate Human Resource Development and Management** | 60.8 | 1,363 |
| 11 | **Higher Diploma HIV and AIDS Management** | 60.2 | 169 |
| 12 | **Certificate HIV and AIDS Management** | 59.7 | 489 |
| 13 | **Higher Certificate Travel, Tourism, and Hospitality** | 58.6 | 681 |
| 14 | **Higher Certificate Finance Management** | 58.4 | 1,538 |
| 15 | **Certificate Business Information Systems** | 58.1 | 909 |
| 16 | **Certificate Travel, Tourism, and Hospitality** | 57.5 | 449 |
| 17 | **Certificate Human Resource Development and Management** | 57.5 | 1,454 |
| 18 | **Honours Degree in HIV and AIDS Management** | 57.4 | 195 |
| 19 | **Honours Degree Digital Communication Technology** | 57.2 | 680 |

| Rank | Degree Programme | Average Final Mark | Number of Final Marks |
|------|------------------|--------------------|-----------------------|
| 20 | Preparatory Courses | 57.0 | 515 |
| 21 | Certificate Finance Management | 56.6 | 1,570 |
| 22 | Honours Degree Business Administration | 56.0 | 549 |
| 23 | Honours Degree Travel, Tourism, and Hospitality | 55.1 | 223 |
| 24 | Honours Degree Marketing and Management | 54.9 | 166 |
| 25 | Honours Degree Finance Management | 54.8 | 715 |
| 26 | Certificate Marketing Management | 53.8 | 323 |
| 27 | Higher Certificate Business Information Systems | 53.5 | 721 |
| 28 | Certificate Business Administration | 53.3 | 1,214 |
| 29 | Higher Diploma Human Resource Development and Management | 52.9 | 465 |
| 30 | Higher Certificate Business Administration | 52.8 | 1,173 |
| 31 | Diploma Secretarial and Computer Studies | 52.3 | 80 |
| 32 | Honours Degree Business Information Systems | 51.7 | 691 |
| 33 | Higher Diploma Travel, Tourism, and Hospitality | 50.5 | 207 |
| 34 | Higher Diploma Marketing Management | 47.7 | 102 |
| 35 | Higher Diploma Finance Management | 45.3 | 390 |
| 36 | Higher Diploma Business Administration | 44.2 | 317 |
| 37 | Higher Diploma Business Information Systems | 43.7 | 432 |

*Data table 4: degree programmes with average final marks, query used "average final marks by degree programme".*

| Source | Sum of Squares | df | Mean Square | F | Sig. |
|---|---|---|---|---|---|
| Degree Programme | 438,977.62 | 42 | 10,451.85 | 49.38 | 0.00 |
| Error | 4.132E+06 | 19,523 | 211.65 | | |
| Total | 4.571E+06 | 19,565 | | | |

*Data table 5: analysis of variance (ANOVA) table for degree programmes. The F is statistically relevant, so reject the null hypothesis that the final mark means are equal across categories of the degree programme.*

**Findings:**

1. With a confidence level of 95.0% in a two-sided significance test, degree programmes in ranks 1 to 9 show significantly higher average final marks than the average of all degree programmes (56.8 out of 100).

2. Degree programmes in ranks 34 to 37 (orange-boxed area in above data table) show average final marks with low academic performance, i.e. below 50 out of 100.

**Association with previous school attendance:**

| Rank | School Code and Name | Average Final Mark | Number of Final Marks |
|---:|---|---:|---:|
| 1 | 7948 Mwadikange Kaulinge Secondary School | **62.8** | 109 |
| 2 | 8276 Iihenda Secondary School | **62.7** | 32 |
| 3 | 7757 Romanus Kamunoko Secondary School | **62.4** | 36 |
| 4 | 7383 Paresis Secondary School | **62.0** | 46 |
| 5 | 8421 Mweshipandeka Senior Secondary School | 61.4 | 118 |
| 6 | 7177 Sangwali Secondary School | 61.4 | 30 |
| 7 | 7008 Swakopmund Secondary School | 61.3 | 30 |
| 8 | 8119 Eengendjo Senior Secondary School | 60.4 | 81 |
| 9 | 7305 Augustineum Secondary School | 60.3 | 113 |
| 10 | 7339 Immanuel Shifidi Secondary School | 60.2 | 111 |
| 11 | 7939 Acacia High School | 60.0 | 40 |
| 12 | 7840 Eldorado Secondary School | 60.0 | 89 |
| 13 | 7362 Otjikoto Senior Secondary School | 59.7 | 66 |
| 14 | 8582 Eheke Senior Secondary School | 59.6 | 52 |
| 15 | 7896 Jan Jonker Afrikaner Secondary School | 59.4 | 95 |
| 16 | 8473 Mwaala High School | 59.3 | 173 |
| 17 | 7228 Kuisebmond Secondary School | 59.2 | 121 |
| 18 | 8303 Hans Daniel Namuhuya Senior Secondary School | 59.1 | 69 |
| 19 | 7419 Mureti Secondary School | 59.0 | 40 |
| 20 | 8602 Haimbili-Haufiku Senior Secondary School | 59.0 | 180 |
| 21 | 7941 Highline Secondary School | 59.0 | 43 |
| 22 | 8246 Ongha Secondary School | 58.8 | 123 |
| 23 | 7043 J. G. van der Wath Junior Secondary School | 58.7 | 30 |
| 24 | 7453 Rundu Secondary School | 58.6 | 106 |
| 25 | 8773 Onguti Senior Secondary School | 58.5 | 96 |
| 26 | 7012 Centaurus Secondary School | 58.5 | 65 |

| Rank | School Code and Name | Average Final Mark | Number of Final Marks |
|---|---|---|---|
| 27 | 7004 Mariental Secondary School | 58.5 | 39 |
| 28 | 7006 Outjo Secondary School | 58.4 | 34 |
| 29 | 7120 St. Kizito College | 58.4 | 60 |
| 30 | 8202 St. Marys Odibo High School | 58.3 | 71 |
| 31 | 7011 Academia Secondary School | 58.3 | 93 |
| 32 | 8262 Ponhofi Secondary School | 58.1 | 177 |
| 33 | 7236 De Duine Secondary School | 58.1 | 33 |
| 34 | 8289 Andimba Toivo ya Toivo Senior Secondary School | 58.0 | 180 |
| 35 | 7835 David Bezuidenhout High School | 58.0 | 81 |
| 36 | 7164 Caprivi Senior Secondary School | 57.9 | 162 |
| 37 | 8753 Ekulo Senior Secondary School | 57.8 | 133 |
| 38 | 7263 Khomas High School | 57.7 | 108 |
| 39 | 8084 Ashipala Junior Secondary School | 57.6 | 77 |
| 40 | 8677 Ruacana Vocational Secondary School | 57.6 | 84 |
| 41 | 8254 Oshela Secondary School | 57.6 | 132 |
| 42 | 7261 Hage G. Geingob High School | 57.5 | 123 |
| 43 | 7000 Wennie du Plessis Senior Secondary School | 57.4 | 709 |
| 44 | 8038 Etalaleko Senior Secondary School | 57.3 | 210 |
| 45 | 7056 Tsumeb Senior Secondary School | 57.3 | 34 |
| 46 | 7010 Etosha Secondary School | 57.3 | 109 |
| 47 | 7390 Jakob Marengo Secondary School | 57.3 | 78 |
| 48 | 8491 Nuuyoma Senior Secondary School | 57.2 | 241 |
| 49 | 7382 A. Shipena Senior Secondary School | 57.2 | 182 |
| 50 | 8762 Negumbo Senior Secondary School | 57.2 | 83 |
| 51 | 8490 Shikongo-Iipinge Senior Secondary School | 57.0 | 105 |
| 52 | 8341 Nehale Senior Secondary School | 56.9 | 164 |
| 53 | 8053 Iipumbu Senior Secondary School | 56.9 | 229 |

| Rank | School Code and Name | Average Final Mark | Number of Final Marks |
|---|---|---|---|
| 54 | **8609 Gabriel Taapopi Senior Secondary School** | 56.7 | 155 |
| 55 | **7311 Concordia College** | 56.0 | 162 |
| 56 | **7001 Grootfontein Senior Secondary School** | 55.8 | 43 |
| 57 | **8680 Uukule Senior Secondary School** | 55.7 | 235 |
| 58 | **7802 Ella Du Plessis Senior Secondary School** | 55.7 | 156 |
| 59 | **7451 Leevi Hakusembe Secondary School** | 55.6 | 44 |
| 60 | **8073 Oshakati Senior Secondary School** | 55.3 | 192 |
| 61 | **7005 Otjiwarongo Secondary School** | 55.2 | 37 |
| 62 | **8032 Shaanika Nashilongo Senior Secondary School** | 55.2 | 194 |
| 63 | **8543 Cosmos High School** | 54.3 | 40 |
| 64 | **8628 Onesi Senior Secondary School** | 54.1 | 104 |
| 65 | **7146 Ngweze Senior Secondary School** | 53.6 | 49 |
| 66 | **7970 David Sheehama Senior Secondary School** | 52.8 | 182 |
| 67 | **8804 Okahandja Secondary School** | 52.3 | 41 |
| 68 | **7994 Haudano Senior Secondary School** | 51.5 | 67 |
| 69 | **7777 Maria Mwengere Secondary School** | 51.4 | 42 |
| 70 | **7003 P. K. De Villiers Secondary School** | **49.7** | 36 |

*Data table 6: schools with average final marks,*
*query used "average final marks by school".* **Limitation: previous attendance on school and student level is only recorded for 25.7% of the students.**

**Findings:**

1. With a confidence level of 95.0% in a two-sided significance test, schools in ranks 1 to 4 show significantly higher average final marks than the average of all schools (57.7 out of 100).

2. School in rank 70 (orange-boxed area in above data table) shows an average final mark with low academic performance, i.e. below 50 out of 100.

**Map representation of data of previous school attendance:**

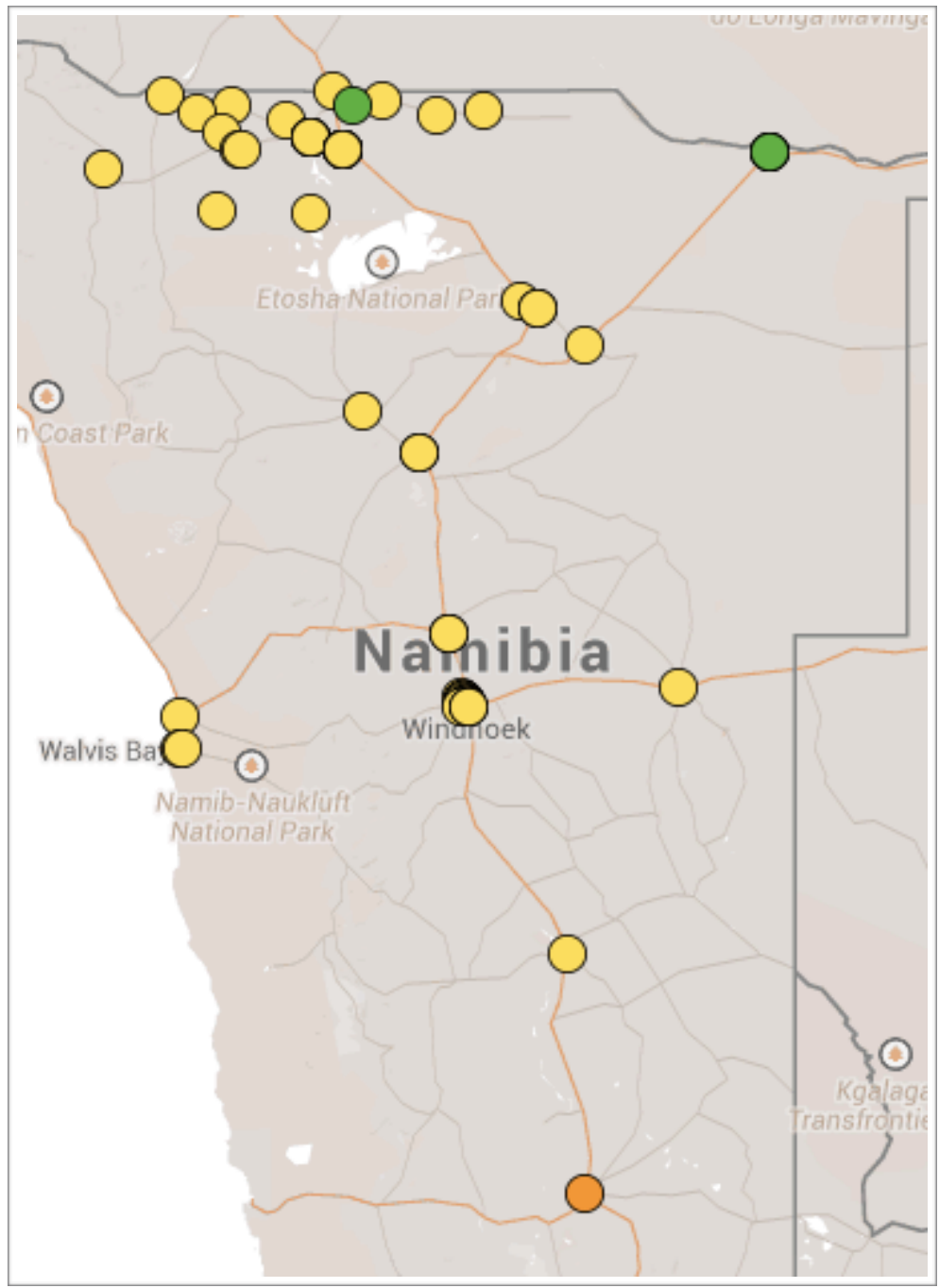

*Figure 4: sample representation of interactive map. Retrieve interactive map from*
*https://www.google.com/maps/d/viewer?mid=z7t3_8C-lGfA.kv4BLgJPN6h0.*
*Green dots represent schools with significantly higher than average*
*academic performance, yellow dots represent schools with average final marks, and red dots*
*represent schools with failing average marks of IUM students.*
***Limitation: previous attendance on school and student level is***
***only recorded for 25.7% of the students population at IUM.***

### 3. Extra-Curricular Learning Offerings and Academic Performance

|  | Pre-Test | Post-Test |
|---|---|---|
| **Number of high performers** | 6 | 5 |
| **Number of low performers** | 11 | 5 |
| **Total number of tests** | 43 | 19 |
| **Average mark** | 57.6 out of 100 | **61.8 out of 100** |
| **Standard deviation of mark** | 13.8 | 17.3 |
| **Percentage of high performers** | 14.0% | **26.3%** |
| **Percentage of low performers** | 25.6% | 26.3% |

*Data table 7: marks by test type used.*

For downloadable, corresponding detail data of the data table see Appendix A, Raw Data Sets.

**Findings:**

1. By offering extra-curricular learning programmes, the average mark between pre- and post-test results shows some improvement, but it is not significantly higher.

2. Percentage of high performing students has increased, but the sample size is too small to derive a significance.

3. Further studies have to be conducted to get a clearer picture.

## 4.  Conclusion

Male students are more likely to be low performers than female students, as well as students born before 1990. Individuals belonging to these two, not necessarily distinct, groups need more guidance and understanding their individual problems to succeed in academic life. Guidance should be offered as an ongoing process, that starts as early as possible in the semester and not just at the end, when most time has passed by. The implementation of summer schools (or rather winter schools in Namibia, i.e. in June and July of each year) could help make up for time lost during the semester.

Some degree programmes at IUM show sub-standard academic performance of students. More supervision and monitoring of lecturers and students by heads of departments, the dean as well as by the quality management of the institution could greatly enhance student performance. Application of accepted quality management frameworks can be of help.

Schools on the lower end of the ranking show sub-standard academic performance of students. The education authorities should closely monitor those schools and their personnel, to identify and rectify problems that lead to low-performing students.

Namibian Universities should offer extra-curricular learning programmes to help build their students' academic knowledge in subjects such as mathematics. This could be done by offering e-learning programmes and digital libraries for e-books. It seems, that the majority of students in Namibia embrace the learning style of accommodation (Kolb, 1974). According to Manochehr (2006), these students are best-served in the form of instructor based e-learning. Students in Namibia need close supervision, guidance and clear goals, to understand the benefits and to undertake the sometimes burdensome journey of additional work.

Students showed great enthusiasm for the e-learning initiative and adapting technical gadgets and functions quickly, this shows digital mentality in Namibia is high. This is further shown by the fact, that Namibia was the first country in Africa to have successfully implemented electronic voting machines (EVM) in a nation-wide election[23]. Considering that 62% of Namibia's population (according to the World Bank, 2010) live in widespread rural areas this shows even more the willingness of the government and the citizens to take steps into a modern, digital society.

And as formerly segregated societies of the world comes closer, one can see a clear distinction of Generation Z in Namibia as it can be observed in western countries. Their main communication product are hand-held devices which can be utilised for modern ways of learning.

| Talking a different language | | | | | |
|---|---|---|---|---|---|
| Formative experiences | **Maturists**<br>(pre-1945)<br>Wartime rationing<br>Rock'n'roll<br>Nuclear families<br>Defined gender roles - particularly for women | **Baby boomers**<br>(1945-1960)<br>Cold War<br>'Swinging Sixties'<br>Moon landings<br>Youth culture<br>Woodstock<br>Family-orientated | **Generation X**<br>(1961-1980)<br>Fall of Berlin Wall<br>Reagan/Gorbachev/Thatcherism<br>Live Aid<br>Early mobile technology<br>Divorce rate rises | **Generation Y**<br>(1981-1995)<br>9/11 terrorists attacks<br>Social media<br>Invasion of Iraq<br>Reality TV<br>Google Earth | **Generation Z**<br>(Born after 1995)<br>Economic downturn<br>Global warming<br>Mobile devices<br>Cloud computing<br>Wiki-leaks |
| Percentage in UK workforce | 3% | 33% | 35% | 29% | Employed in either part-time jobs or apprenticeships |
| Attitude toward career | Jobs for life | Organisational – careers are defined by employees | "Portfolio" careers – loyal to profession, not to employer | Digital entrepreneurs – work "with" organisations | Multitaskers - will move seamlessly between organisations and "pop-up" businesses |
| Signature product | Automobile | Television | Personal computer | Tablet/smartphone | Google glass, 3-D printing |
| Communication media | Formal letter | Telephone | E-mail and text message | Text or social media | Hand-held communication devices |
| Preference when making financial decisions | Face-to-face meetings | Face-to-face ideally but increasingly will go online | Online - would prefer face-to-face if time permitting | Face-to-face | Solutions will be digitally crowd-sourced |

Source: Barclays, University of Liverpool

*Formative social groups. Source: University of Liverpool.*

Based on the findings, one can also conclude, that Namibia looks into a bright future, with students performing better the younger they are, as well as female students, who are showing very promising academic performance. With this in mind, Vision 2030 can be become a reality.

---

[23]  http://www.abc.net.au/news/2014-11-28/namibian-election-first-in-africa-to-use-electronic-voting/5927206

## V.  REFERENCES

Alao, A. (2007). Natural Resources and Conflict in Africa: The Tragedy of Endowment. New York: University of Rochester.

Amutabi, M. N. (2012). Prospects and dilemmas of information and communication technology (ICT) in university education in Africa: The case of Kenya.

Anderson, K. G., Case, A., Lam, D. (2001). Causes and consequences of schooling outcomes in South Africa: Evidence from survey data. Social Dynamics, 27(1), 37-59.

Bandura, A. (1989). Regulation of cognitive processes through perceived self-efficacy. Developmental psychology, 25(5), 729.

Berry, M. V., Brunner, N., Popescu, S., Shukla, P. (2011). Can apparent superluminal neutrino speeds be explained as a quantum weak measurement? J. Phys. A: Math. Theor, 44, 492001.

Bosch, T. E. (2009). Using online social networking for teaching and learning: Facebook use at the University of Cape Town. Communicatio: South African Journal for Communication Theory and Research, 35(2), 185-200.

Brown, A.L., J.D. Bransford, R.A. Ferrara, J.C. Campione (1983). Learning, remembering and understanding, in J. H. Flavell and E. M. Markman (eds.), Handbook of Child Psychology, Cognitive Development. Wiley, 77-166.

Brown, T. (2003). The role of m-learning in the future of e-learning in Africa. In 21st ICDE World Conference. Retrieved from http://www. tml. tkk. fi/Opinnot (Vol. 110).

Campbell, F. A., Ramey, C. T. (1994). Effects of early intervention on intellectual and academic achievement: a follow-up study of children from low-income families. Child development, 65(2), 684-698.

Campbell, M. (1999). Knowledge discovery in deep blue. Communications of the ACM, 42(11), 65-67.

Castro, M., Lopez-Rey, A., Perez-Molina, C. M., Colmenar, A., de Mora, C., Yeves, F., Daniel, J. S. (2001). Examples of distance learning projects in the European Community. Education, IEEE Transactions on, 44(4), 406-411.

Chamberlin, D. D., Gilbert, A. M., & Yost, R. A. (1981, September). A history of system R and SQL/data system. In Proceedings of the seventh international conference on Very Large Data Bases-Volume 7 (pp. 456-464). VLDB Endowment.

Christie, P., Collins, C. (1982). Bantu Education: apartheid ideology or labour reproduction? Comparative Education, 18(1), 59-75.

Clausen, N. (2009). Open Source Business Intelligence. BoD–Books on Demand.

CommonwealthOrg. (2014, December 20). Namibia. Retrieved from The Commonwealth: http://thecommonwealth.org/our-member-countries/namibia

Crook, R. C., Manor, J. (1998). Democracy and decentralisation in South Asia and West Africa: Participation, accountability and performance. Cambridge University Press.

Cross, S. J. (2002). English language proficiency and contextual factors influencing mathematics achievement of secondary school pupils in South Africa. University of Twente.

Darkwa, O., Mazibuko, F. (2000). Creating virtual learning communities in Africa: Challenges and prospects. First Monday, 5(5).

Davis-Kean, P. E. (2005). The influence of parent education and family income on child achievement: the indirect role of parental expectations and the home environment. Journal of family psychology, 19(2), 294.

d'Oliveira, C., Carson, S., James, K., Lazarus, J. (2010). MIT OpenCourseWare: Unlocking knowledge, empowering minds. Science, 329(5991), 525-526.

Edgar, R. R. (2005). The Making of an African Communist: Edwin Thabo Mofutsanyana and the Communist Party of South Africa 1927-1939. Unisa: Unisa Press.

Ellis, J. (1984). Education, repression & liberation: Namibia (Vol. 356). London: World University Service.

Ellison, L. J., Doherty, G. C., Nakos, D., Bhaduri, P., Malla, V., Rossiter, J. (2002). U.S. Patent No. 6,487,547. Washington, DC: U.S. Patent and Trademark Office.

Ellison, L. J., Doherty, C. G., Rossiter, J., Stowell, D., Nakos, D., Bhaduri, P., Sivakumar, T. P. (2004). U.S. Patent No. 6,691,117. Washington, DC: U.S. Patent and Trademark Office.

Ellram, L., Easton, L. (2006). Purchasing Education on the internet. Journal of Management.

Englander, I., Englander, A. (2003). The architecture of computer hardware and systems software: an information technology approach. Wiley.

Entwistle, N. J., Meyer, J. H. F., Tait, H. (1991). Student failure: Disintegrated patterns of study strategies and perceptions of the learning environment. Higher Education, 21(2), 249-261.

Fiske, E. B., Ladd, H. F. (2004). Elusive equity: Education reform in post-apartheid South Africa. Brookings Institution Press.

Fourie, D. J. (1997). Educational language policy and the indigenous languages of Namibia. International journal of the sociology of language, 125(1), 29-42.

Furuholt, B., Ørvik, T. U. (2006). Implementation of information technology in Africa: Understanding and explaining the results of ten years of implementation effort in a Tanzanian organization. Information Technology for Development, 12(1), 45-62.

Gates, B., Myhrvold, N., Rinearson, P. (1995). The road ahead.

Heugh, K. (1999). Languages, development and reconstructing education in South Africa. International journal of educational development, 19(4), 301-313.

Holsinger, D. B., Jacob, W. J. (2009). Inequality in education; comparative and international perspectives. CERC Studies in Comparative Education, 24, 1-33.

Howie, S. (1997). Mathematics and Science Performance in the Middle School Years in South Africa: A Summary Report on the Performance of South African Students in the Third International Mathematics and Science Study (TIMSS).

Howie, S. J. (2003). Language and other background factors affecting secondary pupils' performance in Mathematics in South Africa. African Journal of Research in Mathematics, Science and Technology Education, 7(1), 1-20.

Hyslop, J. (1999). The classroom struggle: Policy and resistance in South Africa, 1940-1990. University of Kwazulu Natal Press.

Jaffer, S., Ng'ambi, D., Czerniewicz, L. (2007). The role of ICTs in higher education in South Africa: One strategy for addressing teaching and learning challenges. International journal of Education and Development using ICT, 3(4).

Khan, S. (2012). The one world schoolhouse: Education reimagined. Hachette Digital.

Kolb, D. (1974). On management and the learning process. Prentice-Hall.

Kamwangamalu, N. M. (2003). Globalization of English, and language maintenance and shift in South Africa. Int'l. J, 165(2516/03), 0164-0065.

Khosrow-Pour, M., Khosrow-Pour, M. (2009). Breakthrough Discoveries in Information Technology Research: Advancing Trends.

Lembede, A. M., Edgar, R. R., Msumza, L. K. (1996). Freedom in our lifetime: the collected writings of Anton Muziwakhe Lembede. Athens: Ohio University Press.

Light, D., Pierson, E. (2014). The Use of Khan Academy in Chilean Classrooms: Study of an Intel Funded Pilot Program in Chile. In Advanced Learning Technologies (ICALT), 2014 IEEE 14th International Conference, 201-203.

Manochehr, N. N. (2006). The influence of learning styles on learners in e-learning environments: An empirical study. Computers in Higher Education Economics Review, 18(1), 10-14.

Marakas, G., Johnson, R., Clay, Paul F. (2007). The Evolving Nature of the Computer Self-Efficacy Construct: An Empirical Investigation of Measurement Construction, Validity, Reliability and Stability Over Time. Journal of the Association for Information Systems: Vol. 8: Iss. 1, Article 2.

Marrades, R., Gutiérrez, Á. (2000). Proofs produced by secondary school students learning geometry in a dynamic computer environment. Educational studies in mathematics, 44(1-2), 87-125.

Miranda H., Amadhila L., Dengeinge R., Shikongo S. (2011). The SACMEQ III Project in Namibia: A Study of the Conditions of Schooling and the Quality of Education.

Möwes, A. D. (2002). The views of educators regarding inclusive education in Namibia (Doctoral dissertation, Stellenbosch: Stellenbosch University).

Mulimila, R. T. (2000). Information technology applications in East Africa government-owned university libraries. Library Review, 49(4), 186-192.

Mwamwenda, T. S. (1994). Gender differences in scores on test anxiety and academic achievement among South African university graduate students. South African Journal of Psychology.

Nitschke, J. J. (2013). Capacity building in trainers of technical vocational education and training at the Namibian College of Open Learning (NAMCOL).

Nkomo, M. O. (1981). The contradictions of Bantu education. Harvard Educational Review, 51(1), 126-138.

Oshikoya, T. W., Hussain, M. N. (1998). Information technology and the challenge of economic development in Africa. African development review, 10(1), 100-133.

Parker, K. (2006). The effect of student characteristics on achievement in introductory microeconomics in South Africa. South African Journal of Economics, 74(1), 137-149.

Pintrich, P. R. (1999). The role of motivation in promoting and sustaining self-regulated learning. International journal of educational research, 31(6), 459-470.

Pretorius, E. J. (2002). Reading ability and academic performance in South Africa: Are we fiddling while Rome is burning?.

Republic of Namibia (2004). Vision 2030.

Republic of Namibia (2012). Namibia's Fourth National Development Plan NDP4.

Scheepers, H., de Villiers, C. (2000). Teaching of a computer literacy course in South Africa: A case study using traditional and co-operative learning. Information Technology for Development, 9(3), 175-187.

Schwalbe, K. (2013). Information technology project management. Cengage Learning.

Setati, M. (2002). Researching mathematics education and language in multilingual South Africa. The Mathematics Educator, 12(2), 6-20.

Spencer, S. J., Steele, C. M., Quinn, D. M. (1999). Stereotype threat and women's math performance. Journal of experimental social psychology, 35(1), 4-28.

Sun, P. C., Tsai, R. J., Finger, G., Chen, Y. Y., Yeh, D. (2008). What drives successful e-learning? An empirical investigation of the critical factors influencing learner satisfaction. Computers & Education, 50(4), 1183-1202.

Sunal, C. S., Mutua, K. (2005). Forefronts in Research (Hc) (Research on Education in Africa, the Caribbean, and the Midd). USA: Information age Publishing Inc.

Thomas, D. (1996). Education across generations in South Africa. The American Economic Review, 330-334.

Turban, E., Leidner, D., McLean, E., Wetherbe, J. (2008). Information technology for management. John Wiley & Sons.

Traxler, J., Dearden, P. (2005). The potential for using SMS to support learning and organisation in Sub-Saharan Africa. In Proceedings of Development Studies Association Conference, Milton Keynes.

UNICEF (2011). Improving quality and equity in education in Namibia. A trend and gap analysis.

Winschiers, H. (2014, December 19). Retrieved from Hamburg State and University Library Carl von Ossietzky (Stabi): http://ediss.sub.uni-hamburg.de/volltexte/2001/482/pdf/Disse.pdf

Wolpe, H. (1972). Capitalism and cheap labour-power in South Africa: from segregation to apartheid 1. Economy and society, 1(4), 425-456.

Zaaiman, H. (1998). Selecting Students for Mathematics and Science: The Challenge Facing Higher Education in South Africa. Human Sciences Research Council, 134 Pretorius Street Pretoria, Private Bag X41, Pretoria, South Africa 0001 (98.02 South African Rand).

Zimmerman, B. J., Schunk, D. H. (Eds.). (2001). Self-regulated learning and academic achievement: Theoretical perspectives. Routledge.

## VI. APPENDIX A

## 1.  SQL Statements to Retrieve Source Data

**Distribution of final marks:**

```
/* distribution of final marks */
SELECT
    TO_NUMBER(I20VACADEMIC_RECORD.FINAL_MARK) AS final_mark
FROM
    STUD.I20VACADEMIC_RECORD
WHERE
    I20VACADEMIC_RECORD.FINAL_MARK IS NOT NULL
ORDER BY
    final_mark DESC;
```

**Distribution of final marks by gender (female):**

```
/* distribution of final marks by gender (female) */
SELECT
    UPPER(IADBIO.IADSEX)                     AS gender,
    TO_NUMBER(I20VACADEMIC_RECORD.FINAL_MARK) AS final_mark
FROM
    STUD.IADBIO,
    STUD.I20VACADEMIC_RECORD
WHERE
    IADBIO.IADSTNO = I20VACADEMIC_RECORD.STUDENT_NUMBER
AND I20VACADEMIC_RECORD.FINAL_MARK IS NOT NULL
AND UPPER(IADBIO.IADSEX) = 'F'
ORDER BY
    final_mark DESC;
```

**Distribution of final marks by gender (male):**

```
/* distribution of final marks by gender (male) */
SELECT
    UPPER(IADBIO.IADSEX)                     AS gender,
    TO_NUMBER(I20VACADEMIC_RECORD.FINAL_MARK) AS final_mark
FROM
    STUD.IADBIO,
    STUD.I20VACADEMIC_RECORD
WHERE
    IADBIO.IADSTNO = I20VACADEMIC_RECORD.STUDENT_NUMBER
AND I20VACADEMIC_RECORD.FINAL_MARK IS NOT NULL
AND UPPER(IADBIO.IADSEX) = 'M'
ORDER BY
    final_mark DESC;
```

**Average final marks by year of birth:**

```
/* average final marks by year of birth */
SELECT
    TO_NUMBER(TO_CHAR(IADBIRDAT,'YYYY')) AS year_of_birth,
    AVG(TO_NUMBER(FINAL_MARK))           AS avg_final_mark,
    COUNT(1)                             AS counter
FROM
    STUD.IADBIO,
    STUD.I20VACADEMIC_RECORD
WHERE
    IADBIO.IADSTNO = I20VACADEMIC_RECORD.STUDENT_NUMBER
AND I20VACADEMIC_RECORD.FINAL_MARK IS NOT NULL
AND TO_NUMBER(TO_CHAR(IADBIO.IADBIRDAT,'YYYY')) > 1914
GROUP BY
    TO_NUMBER(TO_CHAR(IADBIO.IADBIRDAT,'YYYY'))
HAVING
    COUNT(1) >= 30
ORDER BY
    year_of_birth ASC;
```

**Final marks by year of birth:**

```
/* final marks by year of birth */
SELECT
    TO_NUMBER(TO_CHAR(IADBIO.IADBIRDAT,'YYYY')) AS year_of_birth,
    TO_NUMBER(I20VACADEMIC_RECORD.FINAL_MARK)   AS final_mark
FROM
    STUD.IADBIO,
    STUD.I20VACADEMIC_RECORD
WHERE
    IADBIO.IADSTNO = I20VACADEMIC_RECORD.STUDENT_NUMBER
AND I20VACADEMIC_RECORD.FINAL_MARK IS NOT NULL
AND TO_NUMBER(TO_CHAR(IADBIO.IADBIRDAT,'YYYY')) > 1914
ORDER BY
    final_mark DESC;
```

**Distribution of final marks (pre-independence born):**

```
/* distribution of final marks (pre-independence born) */
SELECT
    TO_NUMBER(TO_CHAR(IADBIO.IADBIRDAT,'YYYY')) AS year_of_birth,
    TO_NUMBER(I20VACADEMIC_RECORD.FINAL_MARK)   AS final_mark
FROM
    STUD.IADBIO,
    STUD.I20VACADEMIC_RECORD
WHERE
    IADBIO.IADSTNO = I20VACADEMIC_RECORD.STUDENT_NUMBER
AND I20VACADEMIC_RECORD.FINAL_MARK IS NOT NULL
AND TO_NUMBER(TO_CHAR(IADBIO.IADBIRDAT,'YYYY')) > 1914
AND TO_NUMBER(TO_CHAR(IADBIO.IADBIRDAT,'YYYY')) < 1990
ORDER BY
    final_mark DESC;
```

**Distribution of final marks (post-independence born):**

```
/* distribution of final marks (post-independence born) */
SELECT
    TO_NUMBER(TO_CHAR(IADBIO.IADBIRDAT,'YYYY')) AS year_of_birth,
    TO_NUMBER(I20VACADEMIC_RECORD.FINAL_MARK)  AS final_mark
FROM
    STUD.IADBIO,
    STUD.I20VACADEMIC_RECORD
WHERE
    IADBIO.IADSTNO = I20VACADEMIC_RECORD.STUDENT_NUMBER
AND I20VACADEMIC_RECORD.FINAL_MARK IS NOT NULL
AND TO_NUMBER(TO_CHAR(IADBIO.IADBIRDAT,'YYYY')) >= 1990
ORDER BY
    final_mark DESC;
```

**Average final marks by degree programme:**

```
/* average final marks by degree programme */
SELECT
    I20VACADEMIC_RECORD.QUALIFICATION_NAME AS degr_prog,
    ROUND(AVG(TO_NUMBER(FINAL_MARK)),1)    AS avg_final_mark,
    COUNT(1)                               AS counter
FROM
    STUD.IADBIO,
    STUD.I20VACADEMIC_RECORD
WHERE
    IADBIO.IADSTNO = I20VACADEMIC_RECORD.STUDENT_NUMBER
AND I20VACADEMIC_RECORD.FINAL_MARK IS NOT NULL
GROUP BY
    I20VACADEMIC_RECORD.QUALIFICATION_NAME
HAVING
    COUNT(1) >= 30
ORDER BY
    avg_final_mark DESC;
```

**Average final marks by school:**

```
/* average final marks by school */
SELECT
    IADSCHOOLCODE||' '||IBCNAME AS school,
    ROUND(AVG(FINAL_MARK),1)    AS avg_f_mark,
    COUNT(1)                    AS no_of_marks
FROM
    STUD.IADBIO,
    STUD.I20VACADEMIC_RECORD,
    STUD.IBCSCH
WHERE
    IADBIO.IADSTNO = I20VACADEMIC_RECORD.STUDENT_NUMBER
AND IBCSCH.IBCCODE = IADBIO.IADSCHOOLCODE
AND IADBIO.IADSCHOOLCODE IS NOT NULL
AND I20VACADEMIC_RECORD.FINAL_MARK IS NOT NULL
GROUP BY
    IADSCHOOLCODE||' '||IBCNAME
HAVING
    COUNT(1) >= 30
ORDER BY
    avg_f_mark DESC;
```

## 2. Raw Data Sets

Downloadable at http://goo.gl/gLhxPY.

## 3. Test Material and Other Documents

Downloadable at http://goo.gl/XGnSqR.

## 4. List of Schools in Namibia

Query form at http://www.moe.gov.na/st_li_institutions.php

# VII.APPENDIX B

## 1. Pre-Test S00173

**Research Test I (Pre)**

**Test Date:** _______________________________________

**IUM Student No.:** _____________________________________

**Name:** _______________________________________

E-Mail: _______________________________________

Phone No.: _______________________________________

Birthdate: _______________________________________

Gender:          O male              O female

Degree Program:  _____________  Degree Year:  ___________

I hereby consent to take part in a research study, and I understand I can withdraw participation at any time. No personal data will be published.

**Signature:**        X________________________________________

**Rules:**
1. **Calculators are not allowed.**
2. Books and any other material are not allowed.
3. Scrap paper is ok, but has to be turned in as well.
4. Please do not cheat and talk during the test.
5. **The test consists of 28 questions. You have 60 minutes to complete the test.**
6. **All tests are different, so your neighbour will have different answers than you.**
7. **Circle the 1 (one) correct answer for each question.** The more answers are correct, the better.

S00173

Research Test I (Pre)

IUM Student No.: _________________________________________

**Question 1:**

**The radius of the earth's orbit is 150,000,000,000 meters. What is this number in scientific notation?**

**A**  $1.5 \times 10^{-11}$

**B**  $1.5 \times 10^{11}$

**C**  $15 \times 10^{10}$

**D**  $150 \times 10^{9}$

M00113

**Question 2:**

**Which expression represents 0.0000007 in scientific notation?**

**A**  $7 \times 10^{-9}$

**B**  $7 \times 10^{-7}$

**C**  $7 \times 10^{7}$

**D**  $7 \times 10^{9}$

M2I856

S00173

Research Test I (Pre)

IUM Student No.: _______________________________________

**Question 3:**

$$\frac{11}{12} - \left( \frac{1}{3} + \frac{1}{4} \right) =$$

A $\dfrac{1}{3}$

B $\dfrac{3}{4}$

C $\dfrac{5}{6}$

D $\dfrac{9}{5}$

M02048

**Question 4:**

**One hundred is multiplied by a number between 0 and 1. The answer has to be—**

A less than 0.

B between 0 and 50 but not 25.

C between 0 and 100 but not 50.

D between 0 and 100.

M00275

S00173

IUM Student No.: _______________________________________

**Question 5:**

What is the value of $\left(\dfrac{1}{8}\right)^2$ ?

A  $\dfrac{1}{64}$

B  $\dfrac{1}{32}$

C  $\dfrac{1}{16}$

D  $\dfrac{1}{4}$

M10014

**Question 6:**

**Donald priced six personal Compact Disc (CD) players. The prices are shown below.**

| $21.00, $23.00, $21.00, $39.00, $25.00, $31.00 |

**What is the median price?**

A  $21.00
B  $24.00
C  $27.00
D  $30.00

M02064

S00173

Research Test I (Pre)

IUM Student No.: ______________________________________

**Question 7:**

**The box below shows the number of kilowatt-hours of electricity used last month at each of the houses on Harris Street.**

620, 570, 570, 590, 560, 640, 590, 590, 580

**What is the mode of these data?**

**A** 560

**B** 580

**C** 590

**D** 640

M12248

**Question 8:**

**Three-fourths of the 36 members of a club attended a meeting. Ten of those attending the meeting were female. Which one of the following questions can be answered with the information given?**

**A** How many males are in the club?

**B** How many females are in the club?

**C** How many male members of the club attended the meeting?

**D** How many female members of the club did not attend the meeting?

M00284

S00173

Research Test I (Pre)

IUM Student No.: _______________________________________

**Question 9:**

45. The table below shows the number of real estate transactions by type for a city.

**Real Estate Transactions**

| Type of Property Sold | Number of Sales |
|---|---|
| Single-Family Residence | 157 |
| Condo/Townhouse | 17 |
| Mobile Home | 6 |
| Multi-Family | 2 |
| Commercial | 15 |
| Land | 255 |
| **Total** | **452** |

**Based on the information in the table, which statement is true?**

A  More than half of the sales were single-family residences.

B  More sales occurred for land than in all other areas combined.

C  The number of condo/townhouse sales was more than 10% of the total sales.

D  The number of mobile home and multi-family sales combined was twice the number of commercial sales.

M21305

S00173

Research Test I (Pre)

IUM Student No.: _______________________________________

**Question 10:**

**Which of the following inequalities represents the statement, "A number, $x$, decreased by 13 is less than or equal to 39"?**

**A**   $13 - x \geq 39$

**B**   $13 - x \leq 39$

**C**   $x - 13 \leq 39$

**D**   $x - 13 < 39$

M03049

**Question 11:**

**Divide a number by 5 and add 4 to the result. The answer is 9.**

**Which of the following equations matches these statements?**

**A**   $4 = 9 + \dfrac{n}{5}$

**B**   $\dfrac{n}{5} + 4 = 9$

**C**   $\dfrac{5}{n} = 4$

**D**   $\dfrac{n + 4}{5} = 9$

M00030

S00173

Research Test I (Pre)

IUM Student No.: _______________________________________

**Question 12:**

**Which system of equations represents the statements below?**

The sum of two numbers is ten. One number is five times the other.

A $\begin{cases} xy = 10 \\ y = 5x \end{cases}$

B $\begin{cases} xy = 10 \\ y = x + 5 \end{cases}$

C $\begin{cases} x + y = 10 \\ y = 5x \end{cases}$

D $\begin{cases} x + y = 10 \\ y = x + 5 \end{cases}$

M25231

**Question 13:**

If $h = 3$ and $k = 4,$ then

$$\frac{hk + 4}{2} - 2 =$$

A   6

B   7

C   8

D   10

M00052

S00173

Research Test I (Pre)

IUM Student No.: _______________________________________

**Question 14:**

**After three hours of travel, Car A is about how many kilometers ahead of Car B?**

**A**   2

**B**   10

**C**   20

**D**   25

M00066

S00173

Research Test I (Pre)

IUM Student No.: _________________________________________

**Question 15:**

The graph below compares the weight of an object on Earth to its weight on the Moon.

**An Object's Weight on the Moon**

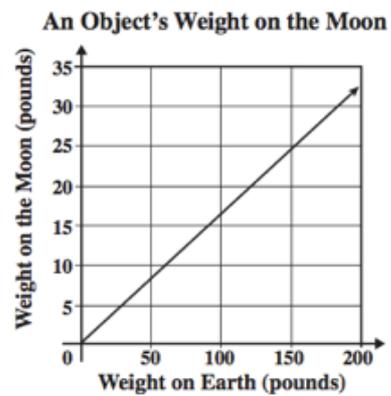

What is the approximate weight on the Moon of an astronaut who weighs 120 pounds on Earth?

A   15 pounds

B   20 pounds

C   25 pounds

D   30 pounds

M10668

Page 10 of 18

S00173

IUM Student No.: _______________________________________

**Question 16:**

**One millimeter is—**

A   $\dfrac{1}{1000}$ of a meter.

B   $\dfrac{1}{100}$ of a meter.

C   100 meters.

D   1000 meters.

M00276

**Question 17:**

**Juanita exercised for one hour. How many seconds did Juanita exercise?**

A      60

B     120

C     360

D   3,600

M03074

S00173

Research Test I (Pre)

IUM Student No.: _______________________________________

**Question 18:**

In Sacramento, the temperature at noon was 35° Celsius (C). What was the temperature in degrees Fahrenheit (F)?

$$\left( F = \frac{9}{5}C + 32 \right)$$

A  35°
B  63°
C  67°
D  95°

M02693

Research Test I (Pre)

IUM Student No.: _______________________________________

**Question 19:**

Javier is using a ruler and a map to
measure the distance from Henley to
Sailport.

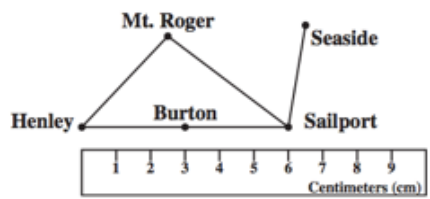

The actual distance from Henley to
Sailport is 120 kilometers (km). What
scale was used to create the map?

**A**  1 cm = 6 km

**B**  1 cm = 12 km

**C**  1 cm = 15 km

**D**  1 cm = 20 km

M21169

**Question 20:**

Marcus can type about 42 words per
minute. If he types at this rate for
30 minutes without stopping, about
how many words will he type?

**A**  1260

**B**  2100

**C**  2520

**D**  4200

M21029

Page 13 of 18

S00173

Research Test I (Pre)

IUM Student No.: _______________________________________

**Question 21:**

A landscaper estimates that
landscaping a new park will take
1 person 48 hours. If 4 people work on
the job and they each work 6-hour days,
how many days are needed to complete
the job?

A  2

B  4

C  6

D  8

M11541

**Question 22:**

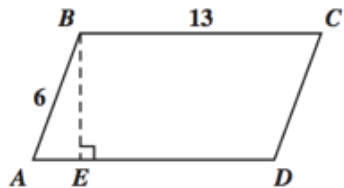

What additional information is needed
to find the area of parallelogram *ABCD*?
$(A = bh)$

A  Length of $\overline{CD}$

B  Length of $\overline{AD}$

C  Length of $\overline{BE}$

D  Perimeter of the parallelogram

M00204

Page 14 of 18

S00173

Research Test I (Pre)

IUM Student No.: _______________________________________

**Question 23:**

The table below shows the flight times from San Francisco (S.F.) to New York (N.Y.).

| Leave S.F. Time | Arrive N.Y. Time |
|---|---|
| 8:30 A.M. | 4:50 P.M. |
| 12:00 noon | 8:25 P.M. |
| 3:30 P.M. | 11:40 P.M. |
| 9:45 P.M. | 5:50 A.M. |

**Which flight takes the longest?**

**A**  The flight leaving at 8:30 A.M.

**B**  The flight leaving at 12:00 noon

**C**  The flight leaving at 3:30 P.M.

**D**  The flight leaving at 9:45 P.M.

M00376

S00173

IUM Student No.: _______________________________________

**Question 24:**

**Use the addition problems below to answer the question.**

$$\frac{1}{2} + \frac{1}{4} = \frac{3}{4}$$

$$\frac{1}{2} + \frac{1}{4} + \frac{1}{8} = \frac{7}{8}$$

$$\frac{1}{2} + \frac{1}{4} + \frac{1}{8} + \frac{1}{16} = \frac{15}{16}$$

$$\frac{1}{2} + \frac{1}{4} + \frac{1}{8} + \frac{1}{16} + \frac{1}{32} = \frac{31}{32}$$

**Based on this pattern, what is the sum of**

$$\frac{1}{2} + \frac{1}{4} + \frac{1}{8} + \frac{1}{16} + \ldots + \frac{1}{1024}?$$

A $\quad \dfrac{1001}{1024}$

B $\quad \dfrac{1010}{1024}$

C $\quad \dfrac{1023}{1024}$

D $\quad \dfrac{1025}{1024}$

M21185

S00173

Research Test I (Pre)

IUM Student No.: ________________________________________

**Question 25:**

If $x$ is an integer, what is the solution
to $|x - 3| < 1$?

A  $\{-3\}$
B  $\{-3, -2, -1, 0, 1\}$
C  $\{3\}$
D  $\{-1, 0, 1, 2, 3\}$

M03035

**Question 26:**

What are all the possible values of $x$
such that $10|x| = 2.5$?

A  $0.25$ and $-0.25$
B  $4$ and $-4$
C  $4.5$ and $-4.5$
D  $25$ and $-25$

M12992

S00173

Research Test I (Pre)

IUM Student No.: ___________________________________

**Question 27:**

**Which of the following is equivalent to $9 - 3x > 4(2x - 1)$?**

**A** $13 < 11x$

**B** $13 > 11x$

**C** $10 > 11x$

**D** $6x > 0$

M02531

**Question 28:**

**Which of the following is equivalent to $1 - 2x > 3(x - 2)$?**

**A** $1 - 2x > 3x - 2$

**B** $1 - 2x > 3x - 5$

**C** $1 - 2x > 3x - 6$

**D** $1 - 2x > 3x - 7$

M02231

**— END OF TEST —**

S00173

## 2.   Pre-Test S00175

Pre-test (document id S00175) shares the same test questions with post-test (document id S00177).

## 3.   Post-Test S00177

**Research Test II (Post)**

**Test Date:** _______________________________________

**IUM Student No.:** _______________________________________

**Name:** _______________________________________

E-Mail: _______________________________________

Phone No.: _______________________________________

Birthdate: _______________________________________

Gender:          O male              O female

Degree Program: _____________ Degree Year: ___________

I hereby consent to take part in a research study, and I understand I can withdraw participation at any time. No personal data will be published.

**Signature:**          X_______________________________________

**Rules:**
1. **Calculators are not allowed.**
2. Books and any other material are not allowed.
3. Scrap paper is ok, but has to be turned in as well.
4. Please do not cheat and talk during the test.
5. **The test consists of 28 questions. You have 60 minutes to complete the test.**
6. **All tests are different, so your neighbour will have different answers than you.**
7. **Circle the 1 (one) correct answer for each question.** The more answers are correct, the better.

S00177

Research Test II (Post)

IUM Student No.: ________________________________________

**Question 1:**

$$3.6 \times 10^2 =$$

**A**    3.600

**B**    36

**C**    360

**D**  3,600

M00036

**Question 2:**

The five members of a band are getting new outfits. Shirts cost $12 each, pants cost $29 each, and boots cost $49 a pair. What is the total cost of the new outfits for all of the members?

**A**    $90

**B**    $95

**C**    $450

**D**    $500

M00331

S00177

Research Test II (Post)

IUM Student No.: ________________________________________

**Question 3:**

**Which of the following numerical expressions results a negative number?**

**A**  $(-7) + (-3)$

**B**  $(-3) + (7)$

**C**  $(3) + (7)$

**D**  $(3) + (-7) + (11)$

M00116

**Question 4:**

**What number equals $\frac{3}{8}$?**

**A**  0.267

**B**  0.375

**C**  2.67

**D**  3.75

M13470

S00177

Research Test II (Post)

IUM Student No.: ___________________________________

**Question 5:**

The cost of an afternoon movie ticket last year was $4.00. This year an afternoon movie ticket costs $5.00. What is the percent increase of the ticket from last year to this year?

A  10%

B  20%

C  25%

D  40%

M02158

**Question 6**

Rico's first three test scores in biology were 65, 90, and 73. What was his mean score?

A  65

B  73

C  76

D  90

M02247

Page 4 of 18

S00177

Research Test II (Post)

IUM Student No.: _______________________________________

**Question 7:**

**The Smithburg town library wanted to see what types of books were borrowed most often.**

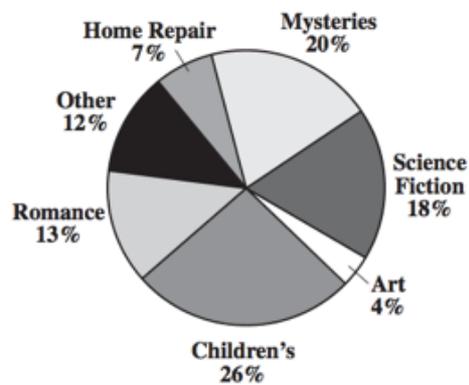

**According to the circle graph shown above—**

**A** more Children's books were borrowed than Romance and Science Fiction combined.

**B** more than half of the books borrowed were Children's, Mysteries, and Art combined.

**C** more Mysteries were borrowed than Art and Science Fiction combined.

**D** more than half of the books borrowed were Romance, Mysteries, and Science Fiction combined.

M02131

S00177

Research Test II (Post)

IUM Student No.: ________________________________________

**Question 8:**

The number of games won over
four years for three teams is shown on
the graph below.

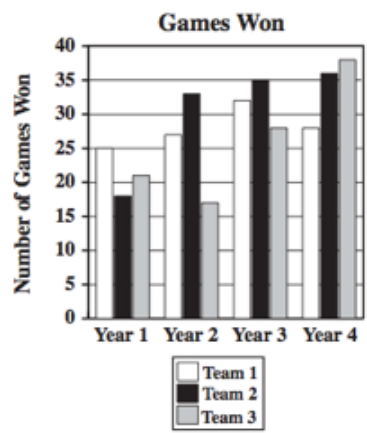

Which statement is true based on this
information?

**A**  Team 3 always came in second.

**B**  Team 1 had the best average overall.

**C**  Team 1 always won more games than
Team 3.

**D**  Team 2 won more games each year
than in the previous year.

M03300

Page 6 of 18

S00177

IUM Student No.: ________________________________________

**Question 9:**

**A student asked 50 children to choose between two colors. The results of the survey are shown in the table below.**

**Color Survey**

| Color | Number |
|-------|--------|
| Pink | 21 |
| Purple | 29 |

**Based on the data in the table, the student claimed that purple is the favorite color of most of the children. Which reason BEST describes why this is an invalid claim?**

**A** Not all of the children chose purple.

**B** More of the children chose pink than purple.

**C** The total number of votes did not equal 50.

**D** The children were only given a choice of two colors.

M32759

S00177

Research Test II (Post)

IUM Student No.: _________________________________________

**Question 10:**

A shopkeeper has $x$ kilograms of tea in stock. He sells 15 kilograms and then receives a new shipment weighing $2y$ kilograms. Which expression represents the weight of the tea he now has?

A   $x - 15 - 2y$

B   $x + 15 + 2y$

C   $x + 15 - 2y$

D   $x - 15 + 2y$

M00110

**Question 11:**

At a local bookstore, books that normally cost $b$ dollars are on sale for 10 dollars off the normal price. How many dollars does it cost to buy 3 books on sale?

A   $3b - 10$

B   $3b + 10$

C   $3(b - 10)$

D   $3(b + 10)$

M10375

S00177

Research Test II (Post)

IUM Student No.: _________________________________________

**Question 12:**

If $n = 2$ and $x = \dfrac{1}{2}$, then
$n(4 - x) =$

A   1
B   3
C   7
D   10

M0003M

**Question 13:**

What is the value of $(3 + 5^2) \div 4 - (x + 1)$
when $x = 7$?

A   −7
B   −1
C   8
D   10

M17663

**Question 14:**

$\sqrt{4x^4} =$

A   2
B   $2x$
C   $4x$
D   $2x^2$

M03067

S00177

Research Test II (Post)

IUM Student No.: _______________________________________

**Question 15:**

**Which expression is equivalent to $7a^2b \cdot 7bc^2$?**

**A**   $14a^2b^2c^2$

**B**   $49a^2bc^2$

**C**   $49a^2b^2c^2$

**D**   $343a^2b^2c^2$

M12872

**Question 16:**

**The actual width ($w$) of a rectangle is 18 centimeters (cm). Use the scale drawing of the rectangle to find the actual length ($l$).**

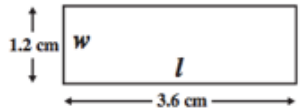

**A**   6 cm

**B**   24 cm

**C**   36 cm

**D**   54 cm

M02087

S00177

Research Test II (Post)

IUM Student No.: _______________________________________

**Question 17:**

Marcus can type about 42 words per
minute. If he types at this rate for
30 minutes without stopping, about
how many words will he type?

A   1260

B   2100

C   2520

D   4200

M20029

**Question 18:**

What is the area of the shaded region in
the figure shown below?

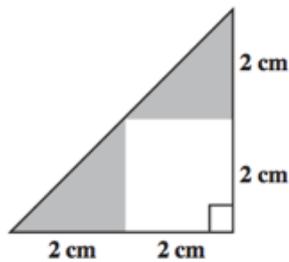

A   4 cm$^2$

B   6 cm$^2$

C   8 cm$^2$

D   16 cm$^2$

M02814

S00177

IUM Student No.: ________________________________________

**Question 19:**

In the figure below, every angle is a right angle.

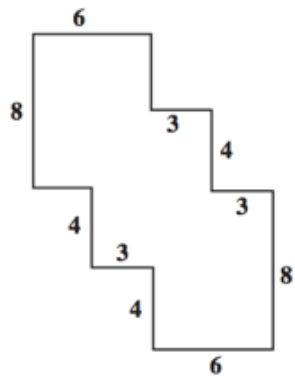

What is the area, in square units, of the figure?

A    96

B    108

C    120

D    144

M10790

Research Test II (Post)

IUM Student No.: _______________________________________

**Question 20:**

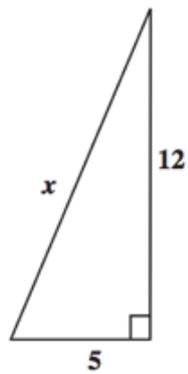

**What is the value of $x$ in the triangle shown above?**

A    11

B    13

C    17

D    169

MD246

S00177

Research Test II (Post)

IUM Student No.: _______________________________________

**Question 21:**

In the diagram below, hexagon *LMNPQR* is congruent to hexagon *STUVWX.*

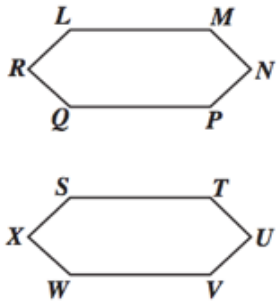

Which side is the same length as $\overline{MN}$ ?

A $\overline{NP}$

B $\overline{TU}$

C $\overline{UV}$

D $\overline{WX}$

M13069

**Question 22:**

If *n* is any odd number, which of the following is true about *n* + 1?

A It is an odd number.

B It is an even number.

C It is a prime number.

D It is the same number as *n* − 1.

M00155

S00177

IUM Student No.: _______________________________________

**Question 23:**

**If *a* is a positive number and *b* is a negative number, which expression is always positive?**

**A** $a - b$

**B** $a + b$

**C** $a \times b$

**D** $a \div b$

M10683

**Question 24:**

**The table below shows the number of visitors to a natural history museum during a 4-day period.**

| Day | Number of Visitors |
|---|---|
| Friday | 597 |
| Saturday | 1115 |
| Sunday | 1346 |
| Monday | 365 |

**Which expression would give the BEST estimate of the total number of visitors during this period?**

**A** $500 + 1100 + 1300 + 300$

**B** $600 + 1100 + 1300 + 300$

**C** $600 + 1100 + 1300 + 400$

**D** $600 + 1100 + 1400 + 400$

M11112

S00177

Research Test II (Post)

IUM Student No.: _______________________________________

**Question 25:**

If $x$ is an integer, which of the following
is the solution set for $3|x| = 15$?

**A** $\{0, 5\}$

**B** $\{-5, 5\}$

**C** $\{-5, 0, 5\}$

**D** $\{0, 45\}$

M00059

**Question 26:**

Which of the following is equivalent
to $4(x+5) - 3(x+2) = 14$?

**A** $4x + 20 - 3x - 6 = 14$

**B** $4x + 5 - 3x + 6 = 14$

**C** $4x + 5 - 3x + 2 = 14$

**D** $4x + 20 - 3x - 2 = 14$

M02936

S00177

Research Test II (Post)

IUM Student No.: _______________________________________

**Question 27:**

$$\frac{20}{x} = \frac{4}{x-5}$$

**Which of the following is equivalent to the equation shown above?**

**A** $x(x-5) = 80$

**B** $20(x-5) = 4x$

**C** $20x = 4(x-5)$

**D** $24 = x + (x-5)$

M02403

S00177

IUM Student No.: _______________________________________

**Question 28:**

**Which equation is equivalent to**
$$\frac{x+3}{8} = \frac{2x-1}{5}?$$

A    $5x + 3 = 16x - 1$

B    $5x + 15 = 16x - 8$

C    $8x + 3 = 10x - 1$

D    $8x + 24 = 10x - 5$

M13117

**— END OF TEST —**

S00177

### 4.   Post-Test S00179

Post-test (document id S00179) shares the same test questions with pre-test (document id S00173).